\documentclass[notitlepage,letterpaper,showpacs,preprintnumbers,amsmath,nofootinbib,amssymb, onecolumn, superscriptaddress, hyperref]{revtex4-1}

\pdfoutput=1
\usepackage{amssymb,amsmath,latexsym,mathrsfs}
\usepackage{graphicx} 
\usepackage{bm}
\usepackage[normalem]{ulem}
\usepackage[usenames,dvipsnames]{color}
\usepackage{hyperref}
\usepackage{comment}
\usepackage{lipsum}
\hypersetup{
    colorlinks=true,       
     citecolor=Blue,
     urlcolor= Blue
    }   
\usepackage{xspace}
\usepackage{soul,xcolor}

\newcommand\be{\begin{equation}}
\newcommand\ee{\end{equation}}
\newcommand\bea{\begin{eqnarray}}
\newcommand\eea{\end{eqnarray}}

\renewcommand\[{\left[}

\begin{document}
\preprint{\tt IFIC/17-10}
\preprint{\tt ULB-TH/17-04}

\title{Warm dark matter and the ionization history of the Universe}

\author{Laura Lopez-Honorez}
\affiliation{Service de Physique Th\'eorique, CP225, Universit\'e Libre de Bruxelles, Bld du Triomphe,
and Theoretische Natuurkunde,
Vrije Universiteit Brussel and The International Solvay Institutes,\\
Pleinlaan 2, B-1050 Brussels, Belgium.}

\author{Olga Mena} 
\affiliation{Instituto de F\'isica Corpuscular (IFIC), CSIC-Universitat de Val\`encia,\\ 
Apartado de Correos 22085,  E-46071, Spain}
\author{Sergio Palomares-Ruiz}
\affiliation{Instituto de F\'isica Corpuscular (IFIC), CSIC-Universitat de Val\`encia,\\ 
Apartado de Correos 22085,  E-46071, Spain}

\author{Pablo Villanueva-Domingo}
\affiliation{Instituto de F\'isica Corpuscular (IFIC), CSIC-Universitat de Val\`encia,\\
Apartado de Correos 22085,  E-46071, Spain}

\begin{abstract}
In warm dark matter scenarios structure formation is suppressed on small scales with respect to the cold dark matter case, reducing the number of low-mass halos and the fraction of ionized gas at high redshifts and thus, delaying reionization. This has an impact on the ionization history of the Universe and measurements of the optical depth to reionization, of the evolution of the global fraction of ionized gas and of the thermal history of the intergalactic medium, can be used to set constraints on the mass of the dark matter particle. However, the suppression of the fraction of ionized medium in these scenarios can be partly compensated by varying other parameters, as the ionization efficiency or the minimum mass for which halos can host star-forming galaxies. Here we use different data sets regarding the ionization and thermal histories of the Universe and, taking into account the degeneracies from several astrophysical parameters, we obtain a lower bound on the mass of thermal warm dark matter candidates of $m_X > 1.3$~keV, or $m_s > 5.5$~keV for the case of sterile neutrinos non-resonantly produced in the early Universe, both at 90\%~confidence level.	
\end{abstract}

\maketitle


\section{Introduction} 

The appearance of the first generation of galaxies, when the Universe was a few hundred million years old, lead to the end of the so-called dark ages of the Universe. The ultraviolet (UV) photons emitted in these galaxies, gradually ionized the neutral hydrogen which had rendered the Universe transparent following the epoch of recombination, in a process known as reionization~\cite{Barkana:2000fd}. However, so far, the exact moment when cosmic reionization took place is not precisely known~\cite{mesinger2016understanding}.

The reionization transition in the late Universe increases the number density of free electrons, $n_e$, which can scatter the Cosmic Microwave Background (CMB), with a probability related to the optical depth at reionization, $\tau$, i.e., the line-of-sight integral of $n_e$ weighted with the Thomson cross section, and dominated by single-ionized hydrogen and helium states. The effect of free electrons on the CMB temperature anisotropies leads to a suppression of the acoustic peaks by a factor $e^{-2\tau}$ at scales within the horizon at the reionization period, a signature which is very degenerate with the amplitude of the primordial power spectrum, $A_s$. Nevertheless, the reionization process creates linear polarization on the CMB spectrum due to the scattering between free electrons and the large-scale CMB quadrupole. This signature, usually dubbed as the ``reionization bump'', with the induced polarized power scaling as $\tau^2$ (see, e.g., Fig.~2 of Ref.~\cite{Reichardt:2015cos}), and peaks at scales larger than the horizon size at the reionization period, resulting in a determination of $\tau$ almost free of degeneracies (see Ref.~\cite{Reichardt:2015cos} or corresponding chapter in Ref.~\cite{mesinger2016understanding} for an exhaustive description of the epoch of reionization (EoR) and its impact on the CMB). Measurements by the Wilkinson Microwave Anisotropy Probe (WMAP) of the optical depth to reionization, $\tau = 0.089 \pm 0.014$, indicated an early-reionization scenario ($z_\mathrm{re} = 10.6 \pm 1.1$)~\cite{Hinshaw:2012aka}, requiring the presence of sources of reionization at $z\gtrsim 10$. This value of $\tau$ was somehow in tension with observations of Lyman-$\alpha$ (Ly-$\alpha$) emitters at $z\simeq 7$~\cite{Stark:2010qj, Treu:2013ida, Pentericci:2014nia, Schenker:2014tda, Tilvi:2014oia}, which instead pointed out to reionization being complete by $z\simeq 6$. Nevertheless, the results presented by the Planck collaboration in the 2015 public data release, including the large-scale (low-$\ell$) polarization observations of the Low Frequency Instrument (LFI)~\cite{Aghanim:2015xee} together with Planck temperature and lensing data, indicate that $\tau = 0.066 \pm 0.016$~\cite{Ade:2015xua} (see also Ref.~\cite{Lattanzi:2016dzq}). Therefore, analyses of the Planck data questioned the need for high-redshift sources of reionization~\cite{Mesinger:2014mqa, Choudhury:2014uba, Robertson:2015uda, Bouwens:2015vha, Mitra:2015yqa}. More recently, an analysis from the Planck collaboration, where unaccounted systematics in the large angular scale polarization data from the High Frequency Instrument (HFI) have been carefully modeled and removed~\cite{Aghanim:2016yuo, Adam:2016hgk}, has provided a measurement of the reionization optical depth of $\tau = 0.055 \pm 0.009$~\cite{Aghanim:2016yuo} based exclusively on the polarization (commonly named as the \emph{EE}) spectrum measurements.

Despite its potential to unravel the mean polarization redshift, the measurement of $\tau$ provides only integrated information on the free electron fraction $x_e$ and not on its precise redshift evolution, i.e., redshift tomography is not possible (see, e.g., Ref.~\cite{Mortonson:2007hq}).  In order to fully characterize such an evolution, upcoming and future measurements of the 21~cm hyperfine transition of neutral hydrogen, which maps its distribution at different redshifts (and thus the distribution of $x_e$), are highly relevant (see, e.g., Refs.~\cite{Furlanetto:2009qk, Pritchard:2011xb, Furlanetto:2015apc, Liu:2015txa}).

Awaiting future cosmological measurements of the $21$~cm transition line, it is important to exploit our current knowledge of the evolution of the total ionized fraction at late times, $\bar{x}_i$. In particular, the redshift dependence of $\bar{x}_i$ allows performing independent and crucial tests of the dark matter (DM) properties, as the free-streaming of light particles at late times could generate differences on the matter power spectrum with respect to the standard cold dark matter (CDM) case. These differences could provide information concerning different particle physics models~\cite{Bode:2000gq, Adhikari:2016bei, Hui:2016ltb}. In this regard, much work has been devoted in the literature to set constraints on annihilations/decays of DM particles from effects on the cosmic ionization history~\cite{Adams:1998nr, Hansen:2003yj, Pierpaoli:2003rz, Chen:2003gz, Padmanabhan:2005es, Zhang:2006fr, Mapelli:2006ej, Shchekinov:2006eb, Furlanetto:2006wp, Valdes:2007cu, Zhang:2007zzh, Chuzhoy:2007fg, Natarajan:2008pk, Cumberbatch:2008rh, Natarajan:2009bm, Belikov:2009qx, Galli:2009zc, Slatyer:2009yq, Huetsi:2009ex, Cirelli:2009bb, Kanzaki:2009hf, Yuan:2009xq, Hisano:2011dc, Galli:2011rz, Finkbeiner:2011dx, Giesen:2012rp, Valdes:2012zv, Slatyer:2012yq, Lopez-Honorez:2013lcm, Galli:2013dna, Diamanti:2013bia, Madhavacheril:2013cna, Evoli:2014pva, Ade:2015xua, Slatyer:2015jla, Kawasaki:2015peu, Lopez-Honorez:2016sur, Liu:2016cnk, Oldengott:2016yjc, Slatyer:2016qyl, Poulin:2016anj}. In this work, on another hand, we focus instead on testing the possibility of warm dark matter (WDM) via measurements of the ionization fraction close to the EoR (see also Refs.~\cite{Lapi:2015zea, Tan:2016xvl}). WDM scenarios offer a very interesting alternative to the standard CDM paradigm, as they provide an explanation of large-scale observations, while alleviating the small-scale problems of the CDM standard picture, i.e., the missing satellite~\cite{Klypin:1999uc, Moore:1999nt}, the too-big-to-fail~\cite{BoylanKolchin:2011dk} and the core-cusp~\cite{Moore:1999gc, Springel:2008cc} problems.~\footnote{It has been argued that some of these problems could possibly be solved once baryonic physics is properly accounted for (see, e.g., Refs.~\cite{Fattahi:2016nld, Lovell:2016nkp}), or via CMB spectral distortions (see the recent Ref.~\cite{Nakama:2017ohe}).} Since WDM particles have non-negligible velocities at high redshifts, structure formation is suppressed at scales below the DM's free-streaming length, delaying the halo and star formation periods. It is precisely this suppression in the growth of small scale structures what allows solving some of the problems of CDM cosmologies mentioned above~\cite{Bode:2000gq, Knebe:2001kb, Colin:2007bk, Zavala:2009ms, Smith:2011ev, Lovell:2011rd, Schneider:2011yu, Lovell:2013ola, Kennedy:2013uta, Destri:2013hha, Angulo:2013sza, Benson:2012su, Kamada:2013sh, Lovell:2015psz, Ludlow:2016ifl, Wang:2016rio}. Furthermore, if the WDM candidate is identified with a keV sterile neutrino, this could provide the origin for the recently observed X-ray signals in galaxy clusters, the galactic center and the cosmic X-ray background~\cite{Bulbul:2014sua, Boyarsky:2014jta, Boyarsky:2014ska, Cappelluti:2017ywp}.

Currently, the most stringent constraints on the WDM particle mass are obtained from the comparison of Ly$\alpha$ forest power spectra observed from distant quasars to the results obtained with hydrodynamical simulations in the non-linear regime, using a power-law temperature-density relation~\cite{Hui:1997dp}. In combination with CMB data, this technique provides lower limits on the WDM mass of  $m_X > 4.17$~keV for a thermal relic\footnote{These particles would decouple early in the Universe and their temperature today should be lower than that of active neutrinos, so that $m_X \, n_X/n_\nu = m_X \, (T_X/T_\nu)^3 = 94.1 \, \Omega_{\rm DM} \, h^2$~eV, where $n_X$ and $n_\nu$ are the number densities of the WDM particle and of active neutrinos and $T_X$ and $T_\nu$ are the corresponding temperatures today.} and $m_s > 25.0$~keV for a non-resonantly produced sterile neutrino, both at $95\%$ confidence level (CL)~\cite{Yeche:2017upn}, improving upon previous bounds~\cite{Viel:2005qj, Seljak:2006qw, Viel:2006kd, Viel:2007mv, Boyarsky:2008xj, Viel:2013apy, Baur:2015jsy}. Even more stringent limits are found when adding the power spectrum measured for $z = 4.2$ and $z = 4.6$ with the high-resolution HIRES/MIKE spectrographs, $m_X > 4.65$~keV (and correspondingly $m_s > 28.8$~keV)~\cite{Yeche:2017upn} and $m_X > 5.3$~keV~\cite{Irsic:2017ixq}. Other bounds obtained from halo counts at high redshifts~\cite{Pacucci:2013jfa, Schultz:2014eia, Dayal:2014nva, Menci:2016eww, Menci:2016eui, Menci:2017nsr}, using the stellar mass function to reproduce the Tully-Fisher relation~\cite{Kang:2012up} or from high-redshift gamma-ray bursts~\cite{deSouza:2013hsj}, are slightly weaker. Let us mention as well that the latest Planck optical depth $\tau$ data has also offered a unique opportunity to improve the existing bounds on reionization driven by a sterile neutrino WDM, produced via resonant oscillations, that could explain the observed X-ray features at $\sim 3.5$~keV (see, e.g., Ref.~\cite{Bose:2016hlz} for a recent and complete analysis of such a possibility). 

In this work, instead, we adopt a different approach and focus on an \emph{universal} aspect of WDM cosmologies, namely, the delay caused in the reionization process due to the free-streaming of DM particles. We shall constrain the WDM scenario via the most recent measurement of $\tau$ from the Planck collaboration~\cite{Aghanim:2016yuo}, together with other constraints on the reionization level at different redshifts. The small-scale suppression of the matter power spectrum, typical in WDM cosmologies, delays structure formation and consequently, the EoR~\cite{Barkana:2001gr, Somerville:2003sh, Yoshida:2003rm}. Exploiting this effect by means of the semi-numerical modeling provided by the {\tt  21cmFAST} code~\cite{Mesinger:2007pd, Mesinger:2010ne}, we shall derive a lower bound on the mass of the WDM particles, devoting special attention to existing degeneracies among the ionization efficiency of UV photons, the number of X-ray photons per solar mass in stars and the minimum virial mass (or temperature) above which halos can start hosting galaxies.

The structure of the paper is as follows. Section~\ref{sec:HMF} presents the modeling of the WDM power spectrum and the resulting halo mass function. Section~\ref{sec:reioWDM} contains the description of the ionization processes and of the different parameters we consider in our analysis, including a discussion of the crucial parameter degeneracies. In Sec.~\ref{sec:data} we describe the data used in our numerical analysis and the results can be found in Sec.~\ref{sec:results}. We draw our conclusions in Sec.~\ref{sec:conclusions}.

\section{Warm dark matter halo mass function}
\label{sec:HMF}

In general, WDM scenarios encompass DM candidates having non-negligible velocities at high redshifts, so the growth of structures is suppressed below a free-streaming length of typically one Mpc. When WDM particles are thermally produced, they are assumed to be relativistic at the epoch of decoupling, in contrast to the standard WIMP scenario, but non-relativistic at the time of matter radiation equality, $t_{\rm eq}$, where substantial growth of perturbations becomes possible. A crude estimate of the free-streaming length can then be obtained computing the distance over which such a particle can travel until $t_{\rm eq}$~\cite{Kolb:1990vq}.  This simplified approach allows one to understand that the free-streaming length decreases with increasing $m_X$ but it misses some important points: the logarithmic growth of perturbations during the radiation dominated era and the fact that free-streaming does not instantaneously switch off after $t_{\rm eq}$. One is thus led to make use of a numerical Boltzmann code so as to accurately account for free-streaming~\cite{Bode:2000gq}. The resulting suppression of the linear matter power spectrum has been fitted,\footnote{Note that the parametrization used in Ref.~\cite{Sitwell:2013fpa} is obtained from Ref.~\cite{Bode:2000gq}.} and can be characterized by~\cite{Viel:2005qj}
\begin{equation}
\label{eq:twdm}
  T_{\rm WDM}(k) = (1+ (\alpha k)^{2\nu})^{-5/\nu} ~, 
\end{equation}
such that the WDM power spectrum can be written in terms of that for CDM as
\begin{equation}
\label{eq:pwdm}
P_{\rm WDM}(k) = P_{\rm CDM}(k) \, T^2_{\rm WDM}(k) ~,
\end{equation}
with $\nu=1.12$ and the breaking scale
\begin{equation}
  \alpha= 0.049 \left(\frac{{\rm keV}}{m_X}\right)^{1.11}\left(\frac{\Omega_X}{0.25}\right)^{0.11}\left(\frac{h}{0.7}\right)^{1.22} \, {\rm Mpc}/h ~,
\end{equation}
where WDM has been assumed to account for all the DM and to be a thermal relic (see also Ref.~\cite{Viel:2005qj} for non-thermal relics).

Around the time of reionization, of interest for this paper, perturbations have long gone non-linear and high-resolution N-body simulations are required to obtain the halo mass function, i.e., the number of halos per unit mass as a function of mass and redshift, which is defined as~\cite{Jenkins:2000bv}
\begin{equation}
\label{eq:dndM}
  \frac{dn(M,z)}{dM} = \frac{\rho_{m,0}}{M^2} \, \frac{d\ln \sigma^{-1}}{d\ln M} \, f(\sigma) ~, 
\end{equation}
where $n(M,z)$ is the (comoving) halo number density, $\rho_{m,0} = \Omega_{m,0} \, \rho_{c,0}$ is the average matter density in the Universe today ($z=0$), $\sigma^2=\sigma^2(M,z)$ is the variance of density perturbations and it is a function of the halo mass $M$ and redshift $z$, and the function $f (\sigma)$ is the first crossing distribution and represents the fraction of mass that has collapsed to form halos per unit interval in $\ln \sigma^{-1}$.  The first analytical derivation of $f(\sigma)$, which is expected to be a universal function, by Press-Schechter, assumed a spherical collapse model. They also used the linear growth of primordial fluctuations to calculate the fraction of mass in virialized objects more massive than a given mass by relating it to the fraction of the volume in which the smoothed initial density field is above some threshold density~\cite{Press:1973iz, Bond:1990iw}. However, within this model, the number of halos is underpredicted for high masses and low redshifts and overpredicted for low masses and redshifts~\cite{Springel:2005nw, Heitmann:2006hr, Lukic:2007fc, Watson:2012mt}. An improvement was achieved by Sheth and Tormen (ST) using the same Press-Schechter formalism but with an ellipsoidal collapse model instead~\cite{Sheth:1999mn, Sheth:1999su, Sheth:2001dp}, resulting in the first crossing distribution $f(\sigma)$ to be given by~\cite{Sheth:1999mn}
\begin{equation}
\label{eq:ST}
f(\sigma) = A \, \sqrt{\frac{2 \, q}{\pi}} \, \left(1 + \left(\frac{\sigma^2}{q \, \delta_{\rm c}^2} \right)^{p} \right)\left(\frac{\delta_{\rm c}}{\sigma}\right) \, e^{-\frac{q \, \delta_{\rm c}^2}{2 \, \sigma^2}} ~,  
\end{equation}
where $q=0.707$ and $p=0.3$ were obtained by fitting the results of the GIF simulations~\cite{Kauffmann:1998gz}, $A=0.322$ is the normalization constant so that $\int f(\sigma) \,  d\ln{\sigma^{-1}} = 1$, and $\delta_{\rm c} = 1.686$ is the critical overdensity required for collapse at $z=0$. Although Ref.~\cite{Sheth:2001dp} later proposed $q=0.75$ in order to reduce the discrepancies with the results of Ref.~\cite{Jenkins:2000bv} at large masses, we will use $q=0.707$ for our default CDM ST halo mass function, following Refs.~\cite{Schneider:2011yu, Schneider:2013ria, Schneider:2014rda}. On the other hand, we use $q=1$ as our default value for WDM scenarios,\footnote{Note that the default values used in {\tt 21cmFAST} are $q = 0.73$, $p = 0.175$ and $A = 0.353$ (also used to describe WDM cosmologies~\cite{Sitwell:2013fpa}), from Ref.~\cite{Jenkins:2000bv}.} value that has been shown to match WDM simulations~\cite{Schneider:2013ria, Schneider:2014rda}. We use the default conditional mass function in {\tt 21cmFAST}, which is based on the hybrid prescription of Refs.~\cite{Barkana:2003qk, Barkana:2007xj}.

In principle, the differences between WDM and CDM scenarios are encoded, via the modification of the matter power spectrum, in the root-mean-square (rms) variance of density perturbations, which is defined as
\begin{equation}
\label{eq:sig2}
\sigma^2(M(R), z) = \left(\frac{D(z)}{D(0)}\right)^{2} \int \frac{d^3k}{(2\pi)^3} \, P(k) \, |W(kR)|^2 ~, 
\end{equation}
where the redshift dependence is driven by the linear growth function, $D(z)$, $P(k)$ is the linear power spectrum at $z=0$ computed following Eq.~(\ref{eq:pwdm}) for WDM and $W(kR)$ is the Fourier transform of a filter function. For CDM the filter function is usually taken to be a spherical top-hat (TH) function in real space, on a scale $R^3=3 \, M/(4\pi \rho_{\rm m,0})$, i.e., 
\begin{equation}
  W_{\rm TH}(kR)= \frac{3}{kR} \, \left(\sin(kR)-3\cos(kR)\right) ~.
  \label{eq:TH}
\end{equation}
Although it has been used previously~\cite{Sitwell:2013fpa}, this appears to be inadequate to describe WDM cosmologies~\cite{Barkana:2001gr, Benson:2012su, Schneider:2013ria}, for which there is a cutoff in the matter power spectrum at small masses. With this choice of filter, the halo mass function increases with decreasing mass, contrary to what is found in WDM simulations. This can be understood by the fact that for a given scale $R$ a large range of unsuppressed scales $k$ contributes to $\sigma^2$ and hence to the halo mass function~\cite{Schewtschenko:2014fca}. Instead, it has been argued that the redshift evolution of the WDM suppression of power at small scales observed in N-body simulations is better accounted for by using a sharp-$k$ window, i.e., a spherical top-hat window in $k$-space~\cite{Benson:2012su, Schneider:2013ria},
\begin{equation}
  \label{eq:SK}
  W_{\rm SK}(kR)= \Theta(1 - kR) ~, 
\end{equation}
where $\Theta$ is the Heaviside function. With this choice of window function, the halo mass function can be written as
\begin{equation}
\label{eq:HMFsk}
\frac{d n_{\rm SK}}{d M_{\rm SK}} = \frac{1}{2} \, \frac{\rho_{m,0}}{M_{\rm SK}^2} \, f(\sigma_{\rm SK}) \, \frac{1}{2 \, \pi^2 \, \sigma_{\rm SK}^2} \, \frac{P(1/R_{\rm SK})}{R_{\rm SK}^3} \, \frac{d \ln R_{\rm SK}}{d \ln M_{\rm SK}} ~,
\end{equation}
where, in this work, $f$ is given by Eq.~(\ref{eq:ST}). In the case, of the spherical top-hat filter in real space, the mass assignment for each scale $R_{\rm SK}$ is unambiguously defined. In contrast, in the case of the sharp-$k$ filter, the mass is not well defined given a scale in real space and needs to be constrained from the results of simulations. Except from the dependence $M_{\rm SK} \propto R_{\rm SK}^3$, guaranteed by the spherical symmetry of the filter, a free parameter $c$ has to be introduced such that
\begin{equation}
  \label{eq:MSK}
  M_{\rm SK} = \frac{4\pi}{3} \rho_m \, (c\, R_{\rm SK})^3 ~.
\end{equation}
In this work, we take $c=2.5$ to match the halo mass functions obtained from simulations~\cite{Benson:2012su, Schneider:2014rda}. In the following, we will thus us a TH filter for CDM and a SK filter for WDM, and we have accordingly modified the {\tt 21cmFAST} code.\footnote{We stress again that to describe the suppression at small scales in WDM scenarios, only the window function (and the power spectrum), but not the first crossing distribution, is different from the standard Sheth-Tormen approach in CDM scenarios. We also stress that the modifications to the code described above are slightly different from those introduced in Ref.~\cite{Sitwell:2013fpa}.}

\begin{figure*}[t]
\begin{center}
\includegraphics[width=.49\textwidth]{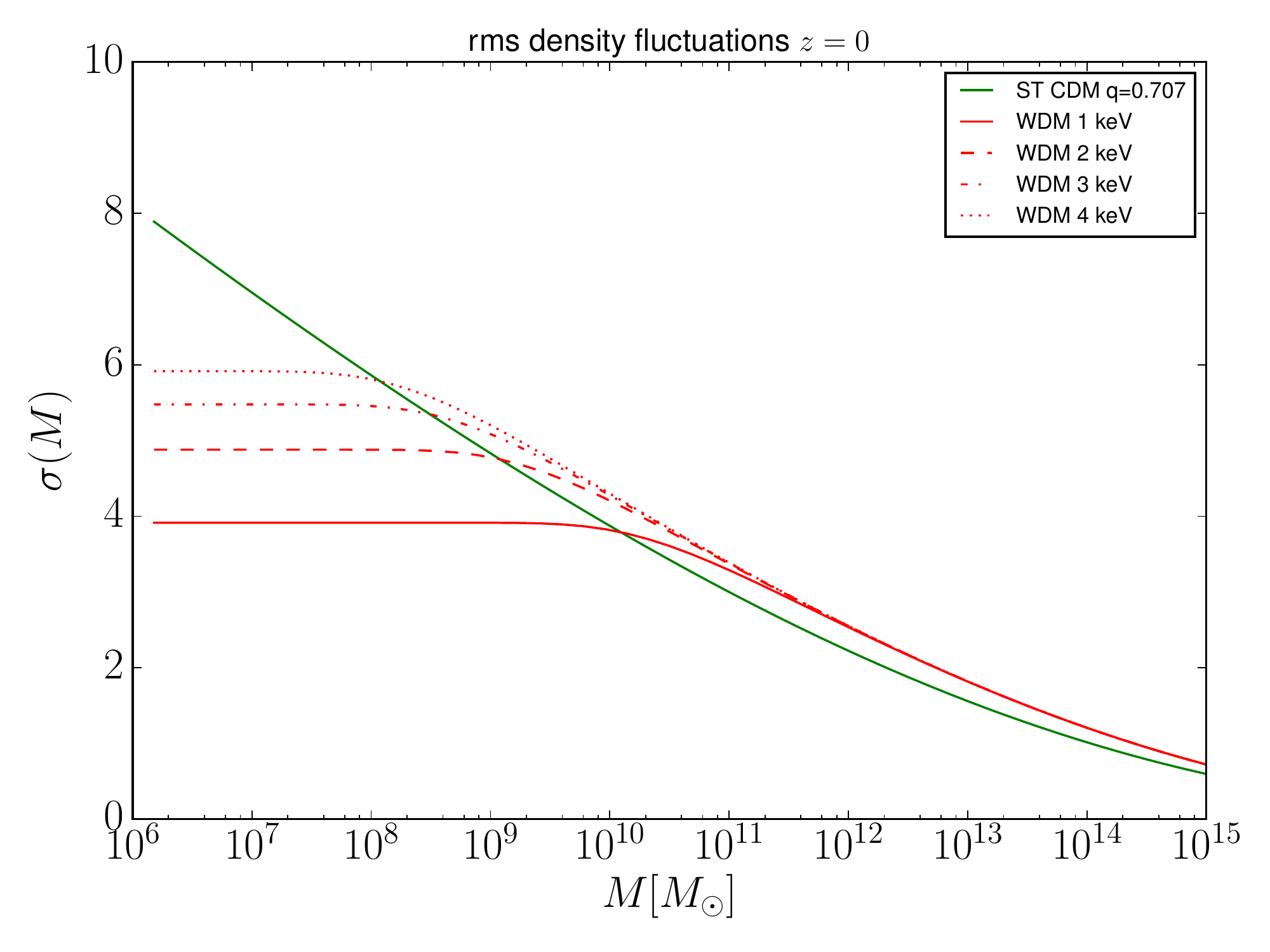} \includegraphics[width=.49\textwidth]{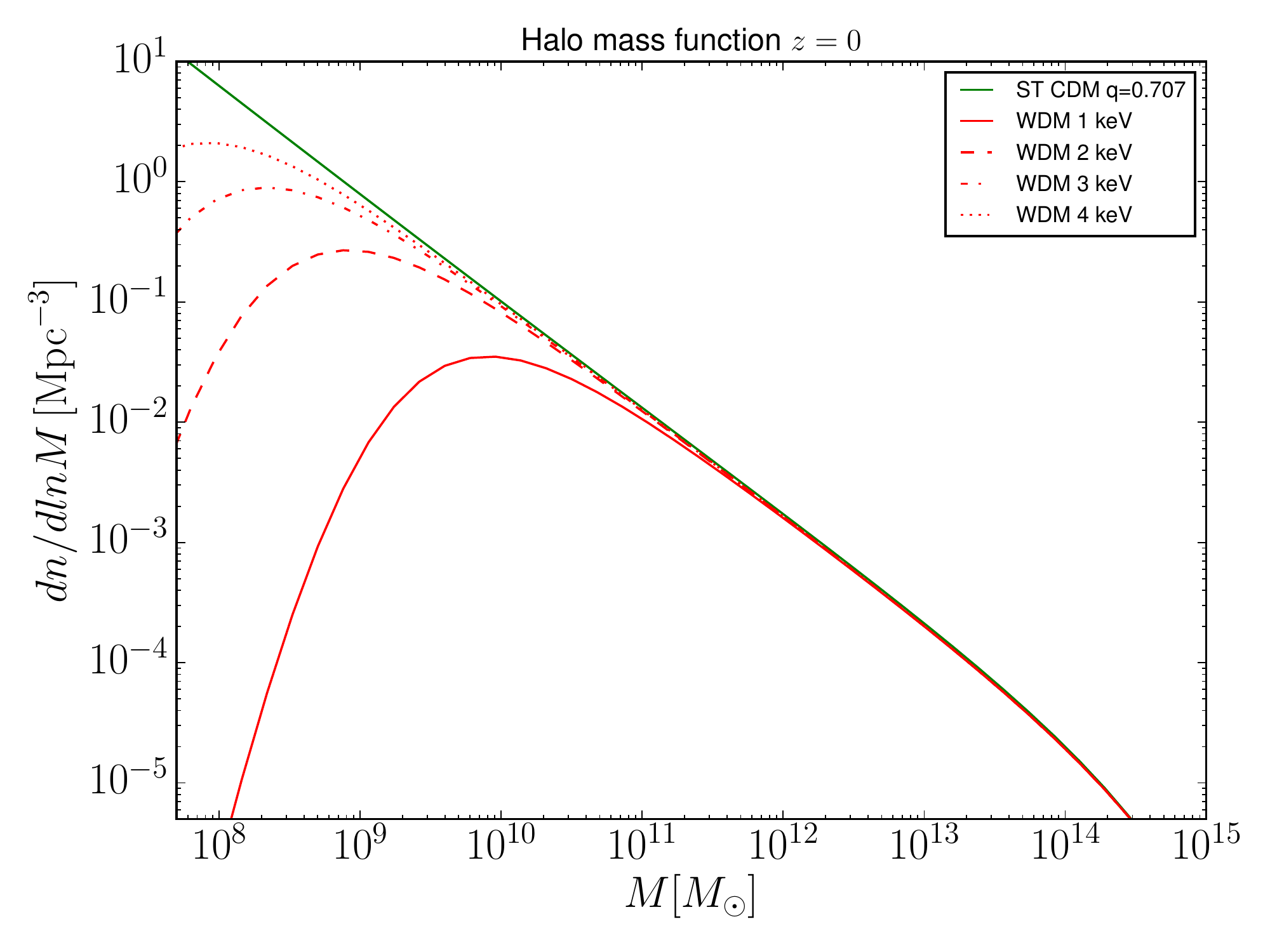} \\  \includegraphics[width=.49\textwidth]{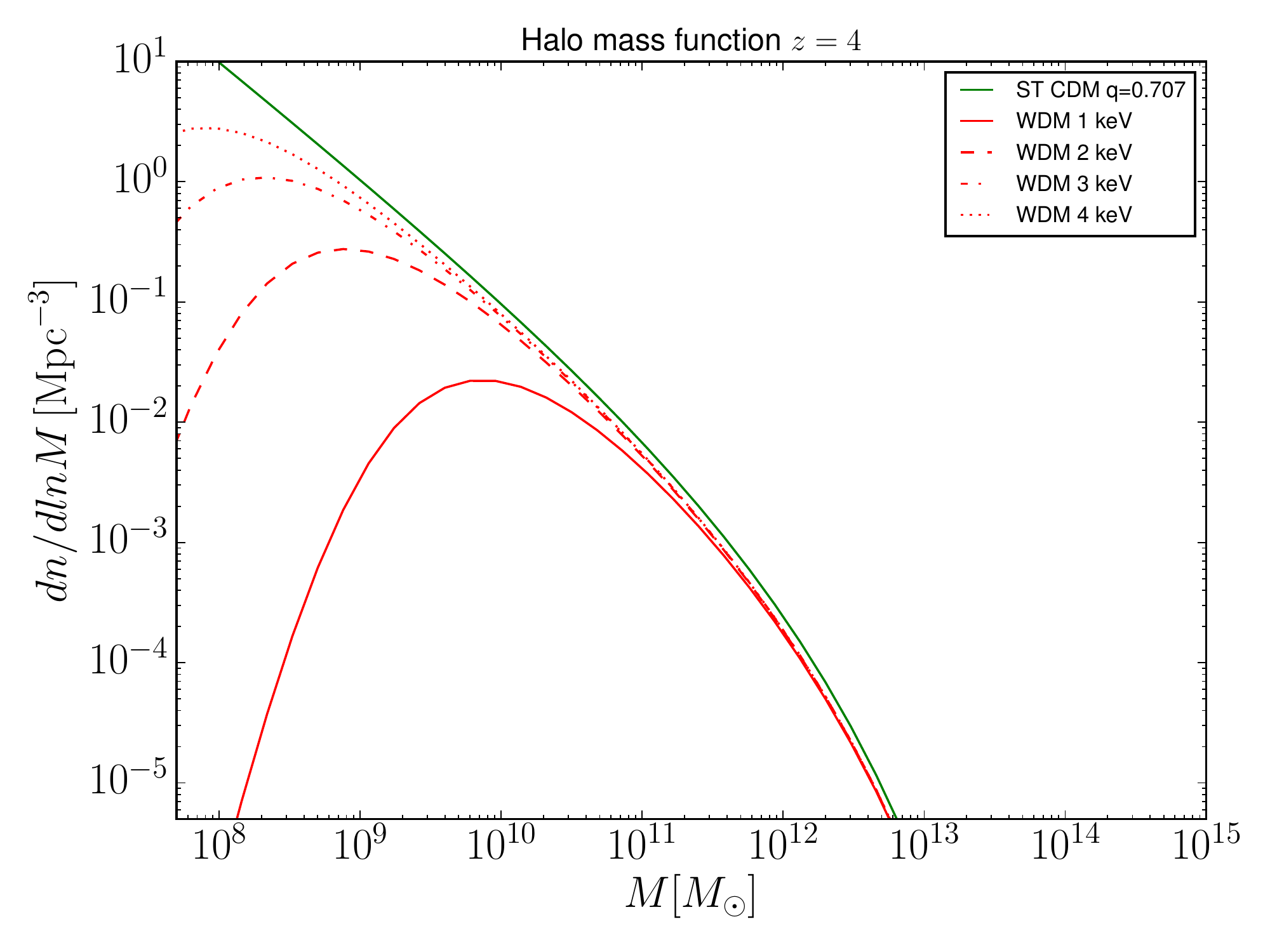} \includegraphics[width=.49\textwidth]{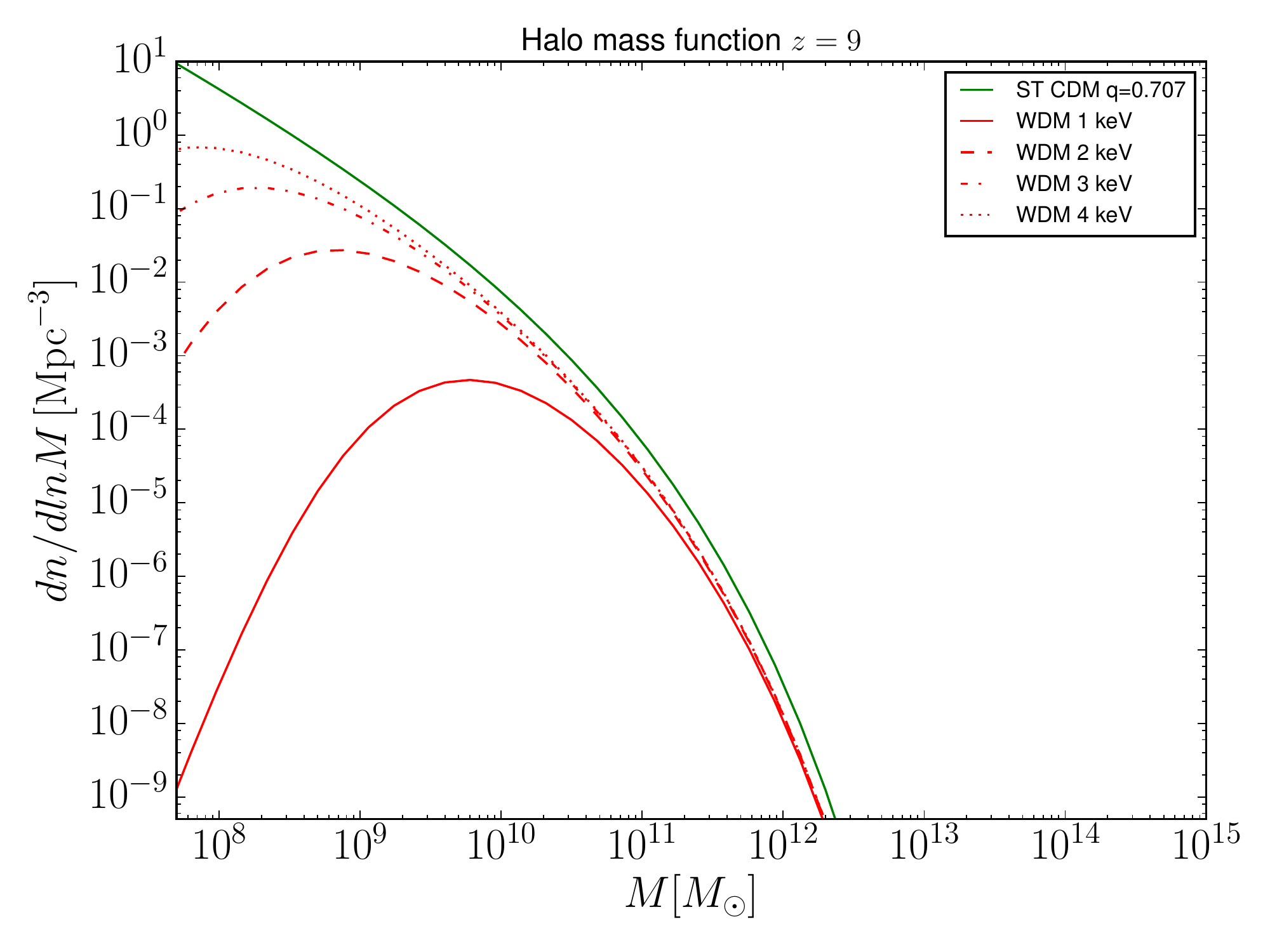} \\
\end{center}
\caption{Root-mean-square density fluctuation $\sigma(M)$ at $z=0$ (top-left panel). Halo mass functions for CDM and WDM at $z=0$ (top-right panel), $z=4$ (bottom-left panel) and $z=9$ (bottom-right panel). For CDM we use $q=0.707$, see Eq.~(\ref{eq:ST}), and a spherical top-hat filter in real space. For WDM, we use $q=1$ and a spherical top-hat filter in $k$-space.}
\label{fig:HMF}
\end{figure*}

Figure~\ref{fig:HMF} depicts the mass dependence of the rms of matter density fluctuations, $\sigma(M)$, and of the halo mass function in WDM scenarios for the range of WDM masses considered in this work. We compare the results for the CDM case with $q=0.707$ and a TH filter to those corresponding to WDM scenarios with $q=1$ and a SK filter. With a TH window function used to describe the CDM case, $\sigma_{\rm TH}(M_{\rm TH})$ increases monotonically with decreasing values of $M_{\rm TH}$.  In contrast, for the WDM SK window function, the rms variance increases monotonically for large masses, while it becomes constant for small masses to account for the free-streaming effects. This is illustrated in the top-left panel of Fig.~\ref{fig:HMF}. The transition between these two regimes is dictated by the abruptness of the cutoff in the linear WDM power spectrum. Notice that, although the shape of $\sigma(M)$ at large masses is similar for WDM and CDM scenarios, their values differ. One should thus  take this into account when normalizing the results obtained from {\tt  21cmFAST}~\cite{Mesinger:2007pd, Mesinger:2010ne}.\footnote{For the best fit of the cosmological parameters, $\sigma_8(z=0)|_{\rm TH}=0.82$ using the TH filter in real space, while $\sigma_8(z=0)|_{\rm SK}=0.48$ using the SK filter.}  The other three panels of Fig.~\ref{fig:HMF} show the resulting halo mass function at redshift $z=0$ (top-right panel), $z=4$ (bottom-left panel) and $z=9$ (bottom-right panel). The flattening of $\sigma(M)$ for low halo masses directly induces the suppression of the halo mass function in the WDM cases. For large halo masses, a similar mass dependence for both WDM and CDM at $z=0$ is found. Let us mention, as reported in Refs.~\cite{Schneider:2013ria, Schneider:2014rda}, that the sharp$-k$ model typically underestimates the halo abundance at large halo masses especially at large redshifts, even though the discrepancy with the data is greatly reduced for halos defined by a spherical overdensity halo finder algorithm. Thus, the halo mass function model for WDM used here does not exactly reproduce the CDM behavior for large halo masses.

Notice that the model for WDM halo mass function of Ref.~\cite{Schneider:2014rda}, which is the one we follow, is very similar to the one used in Ref.~\cite{Benson:2012su} except from one point. Based on Ref.~\cite{Barkana:2001gr}, the authors of Ref.~\cite{Benson:2012su} argue that, due to the WDM residual thermal velocity dispersion at the time of collapse, the growth of collapsing overdensities is suppressed below an effective Jeans mass $M_J$. Below that mass, the critical overdensity for collapse is larger than in CDM scenarios, so the halo mass function is further suppressed. As mentioned in Refs.~\cite{Schneider:2013ria, Schneider:2014rda}, such a Jeans mass is however expected to drop significantly once the Universe enters the matter-dominated era and to damp perturbations on   scales much lower than the free-streaming length. However, this is already accounted for in the above treatment of the WDM case. Thus, in our treatment of the WDM halo mass function, we do not include the effect of late time velocity contributions and we have modified accordingly the {\tt 21cmFAST} code to ensure that no extra Jeans mass cutoff at a given $M_{J}$ has been imposed. As a final note, we stress that the halo mass function we use provides a good description of the outcome from numerical simulations up to redshift $z=5$~\cite{Schneider:2014rda}, even if significant uncertainties are still present. At higher redshifts we have to rely on extrapolations.

\pagebreak
\section{Ionization and thermal histories and warm dark matter}
\label{sec:reioWDM}

As mentioned in the introduction, our goal is to constrain WDM models evaluating the impact on the ionization history of the Universe. For this, we study the evolution of the total ionized fraction $\bar{x}_i$. For that purpose, we make use of the publicly available code {\tt 21cmFast} that, based on excursion set formalism, perturbation theory and analytic prescriptions, generates semi-analytic simulations of the evolved density, peculiar velocity, halo and ionization fields. This code has specially been developed with the purpose of studying variations in the 21 cm signal due to the change in a given set of astrophysical and cosmological parameters. Here we use this code in order to evaluate the ionization fraction evolution around (before) the time of reionization. Notice that $\bar{x}_i ({\bf x}, z)$ is obtained from two separate contributions. In the {\it ionized} IGM (once first sources have lighted on) the ionization level is characterized by $Q_{\rm HII} = \zeta_{\rm UV}f_{\rm coll}(>M_{\rm vir}^{\rm min})$ where $Q_{\rm HII}$ denotes the covering factor of the fully ionized HII regions, $\zeta_{\rm UV}$ characterizes the UV ionization efficiency (see below) and $f_{\rm coll}(>M_{\rm vir}^{\rm min})$ is the fraction of mass collapsed into halos with mass large enough ($>M_{\rm vir}^{\rm min}$) to host star-forming galaxies. The latter is defined in terms the halo mass function introduced in the previous section as
\begin{equation}
\label{eq:fcol}
 f_{\rm coll} (>M_{\rm vir}^{\rm min}) = \int_{M_{\rm vir}^{\rm min}} \frac{M}{\rho_{m,0}} \, \frac{dn}{dM} \, dM ~.
\end{equation}
On the other hand, the local ionized fraction of the \emph{neutral} IGM, $x_e({\bf x}, z)$ can be written as
\begin{equation}
\label{eq:xe}
\frac{dx_e ({\bf x}, z)}{dz}  =  \frac{dt}{dz} \left(\Lambda_{\rm ion}
 - \alpha_{\rm A} \, C \, x_e^2 \, n_b \, \mathfrak{f}_{\rm H} \right) ~,  
\end{equation}
where $n_b=\bar{n}_{b, 0} (1+z)^3 (1+\bar{\delta_b}({\bf x}, z))$ is the baryon number density,  $\Lambda_{\rm ion}$ the ionization rate, $\alpha_{\rm A}$ the case-A recombination coefficient,\footnote{Case-A recombination involves a sum over all recombination coefficients (including recombinations to the ground state) and it is typically used for optically thin media. In this case, one expects that all produced recombination photons escape the system without giving rise to ionization. Case-A is appropriate in the highly-ionized low-redshift Universe, where most of the recombinations are actually taking place in dense, partially neutral gas, so-called Lyman-limit systems (LLS). In this case, photons resulting from ground state recombinations are likely to be absorbed locally, inside the LLS and do not contribute to the ionization balance in the diffuse IGM~\cite{MiraldaEscude:2002yd}. However, notice that considering case-A or case-B recombination has a negligible impact on the results presented here.} $C\equiv \langle n_e^2 \rangle / \langle n_e \rangle^2$ is the clumping factor, with $n_e$ the electron number density, and $\mathfrak{f}_{\rm H}=n_{\rm H}/n_b$ is the hydrogen number fraction. Eq.~(\ref{eq:xe}) is solved numerically by means of the {\tt 21cmFAST} code, briefly described above.

The total ionized fraction reads (see, e.g., Ref.~\cite{Mesinger:2012ys})
\begin{equation}
\label{eq:xetot}
 \bar{x}_i \simeq Q_{\rm HII}+(1-Q_{\rm HII}) \, x_e ~.
\end{equation}
Notice that for the purpose of this work the most relevant contribution to $\bar{x}_i$ is $Q_{\rm HII}$ that drives the ionization fraction around the reionization time. Once the ionization history is at hand, one can compute the optical depth to reionization, defined as
\begin{equation}
\label{eq:tau} 
\tau=\sigma_T\int \tilde{x}_i \, n_b \, dl ~,
\end{equation}
where $\sigma_T$ is the Thomson cross-section and $dl$ is the line-of-sight proper distance.

\subsection{Free astrophysics parameters}

For the sake of simplicity and for comparison purposes with other works, we shall describe the ionization history of the IGM in terms of a reduced number of quantities, namely the WDM mass $m_X$, the ionization efficiency of UV photons $\zeta_{\rm UV}$, the minimum virial temperature $T_{\rm vir}^{\rm min}$ (or equivalently the minimum virial mass $M_{\rm vir}^{\rm min}$, see below) and the X-ray efficiency $\zeta_{\rm X}$. In order to obtain the evolution of $\bar{x}_i (z)$ we make use of the publicly available code {\tt 21cmFAST}~\cite{Mesinger:2007pd, Mesinger:2010ne}~\footnote{Except for the astrophysics parameters mentioned here, we use the {\tt 21cmFAST} default settings for our simulations (for the version we use, we have a (200 Mpc)$^3$ comoving box with a 900$^3$ grid).}. Concerning the range considered for the WDM particle mass in the case of thermal relics, we restrict ourselves to the few keV region, $m_X \in [1-4]$~keV, where previous related analyses have been focused on.

As mentioned above, the UV ionizing efficency $\zeta_{\rm UV}$ fixes the ionization fraction in the ionized IGM. It can be reexpressed in terms of  the fraction of ionizing photons escaping their host galaxy $f_{\rm esc}$, the number of ionizing photons per stellar baryons inside stars $N_\gamma$, the fraction of baryons that form stars $f_\star$, and the mean number of recombinations per baryon $\bar{n}_{\rm rec}$,\footnote{In WDM models, the rate of recombinations in the smallest halos is smaller than in CDM models, which could make reionization to occur earlier than in CDM models. Nevertheless, note that $\bar{n}_{\rm rec}$ is included in our definition of $\zeta_{\rm UV}$, although the trading-off of these parameters is, in general, non-trivial~\cite{Yue:2012na, Rudakovskiy:2016ngi}.} as~\cite{Mesinger:2012ys}
\begin{equation}
\label{eq:zUV}
\zeta_{\rm UV}\simeq 30\, \left(\frac{N_\gamma}{4400}\right) \left(\frac{f_{\rm esc}}{0.1}\right) \left(\frac{f_\star}{0.1}\right) \left(\frac{1.5}{1+\bar{n}_{\rm rec}}\right) ~.
\end{equation}
As already noted in Ref.~\cite{Sitwell:2013fpa}, suppressing the photon-production efficiency can have similar effects on ionization observables, as for WDM the abundance of low mass halos is suppressed. We allow $\zeta_{\rm UV}$ to vary in the range $\zeta_{\rm UV} \in [5,105]$ (see, e.g., Ref.~\cite{Liu:2015txa} for the bounds that could be obtained on this parameter from future 21~cm observations).  Let us emphasize that for the purpose of this work, we have not modified the criterium encoded in {\tt 21cmFast} for a region to be considered ionized~\footnote{ Notice that~\cite{Paranjape:2015ocl} pointed out that the excursion set formalism, used in {\tt 21cmFast}, tracks the average collapsed mass fraction $f_{\rm coll}$ instead of the stochastic source count which can give rise to a non-conservation of photons.} namely:
\begin{equation}
    \zeta_{\rm UV} f_{\rm coll}>1  \,,
\end{equation}
with $\zeta_{UV}$ assumed to be constant with redshift for simplicity.  

Another parameter we allow to vary is $T_{\rm vir}^{\rm min}$,\footnote{ We take the same threshold temperature $T_{\rm vir}^{\rm min}$ for halos hosting ionizing and X-ray sources.} which is the threshold temperature for halos hosting star-forming galaxies. The default value in the numerical code {\tt 21cmFAST} is $T_{\rm vir}^{\rm min} = 10^4$~K, as lower temperatures have been shown to be insufficient to efficiently cool the halo gas through atomic cooling~\cite{Evrard:1990fu, Blanchard:1992, Tegmark:1996yt, Haiman:1999mn, Ciardi:1999mx}. The choice of $T_{\rm vir}^{\rm min}$ can be translated into a minimum virial halo mass value~\cite{Barkana:2000fd}
 \begin{equation}
\label{eq:mminT}
M_{\rm vir}^{\rm min} (z) \simeq 10^8 \left(\frac{T_{\rm vir}^{\rm min}}{2 \times 10^4 \, {\rm K}} \right)^{3/2} \left(\frac{1+z}{10}\right)^{-3/2} M_\odot ~,
 \end{equation} 
which implies, e.g., $M_{\rm vir}^{\rm min} \simeq 3 \times 10^{7} \, M_\odot$ at a redshift $z = 10$ for $T_{\rm vir}^{\rm min} = 10^4$ K. Following the upper limit $T_{\rm vir}^{\rm min} \sim 2 \times 10^5$~K quoted in Refs.~\cite{Mesinger:2012ys, Greig:2015qca}, we shall restrict ourselves to the range $T_{\rm vir}^{\rm min} \in [10^4 - 10^5]$~K.

Finally, we also vary modestly the number of X-ray photons per solar mass in stars, dubbed as $\zeta_{\rm X}$ (see, e.g., Refs.~\cite{Mesinger:2010ne, Christian:2013gma}).  We consider two values: $\zeta_{\rm X} = 10^{56} \, M_\odot^{-1}$ and $5\times 10^{56} \, M_\odot^{-1}$, which approximately correspond to $N_{\rm X} \simeq 0.1$ and 0.5 X-ray photons per stellar baryon. Although this range is consistent with the observed integrated $0.5-8$~keV luminosity at $z=0$~\cite{Mineo:2011id}, we note that significant uncertainties exist~\cite{Treister:2011xr, Cowie:2011qd, BasuZych:2012tx}. Nevertheless, given the degeneracies in the current analysis,\footnote{In the case of $\bar{x}_i (z)$, a value of $\zeta_{\rm X}$ larger than the range considered here can be approximately traded off for a larger value of $T_{\rm vir}^{\rm min}$ or a lower value of $\zeta_{\rm UV}$.} we restrict $\zeta_{\rm X}$ to that limited range.

\begin{figure*}[t]
	\begin{center}
		\includegraphics[width=0.49\textwidth]{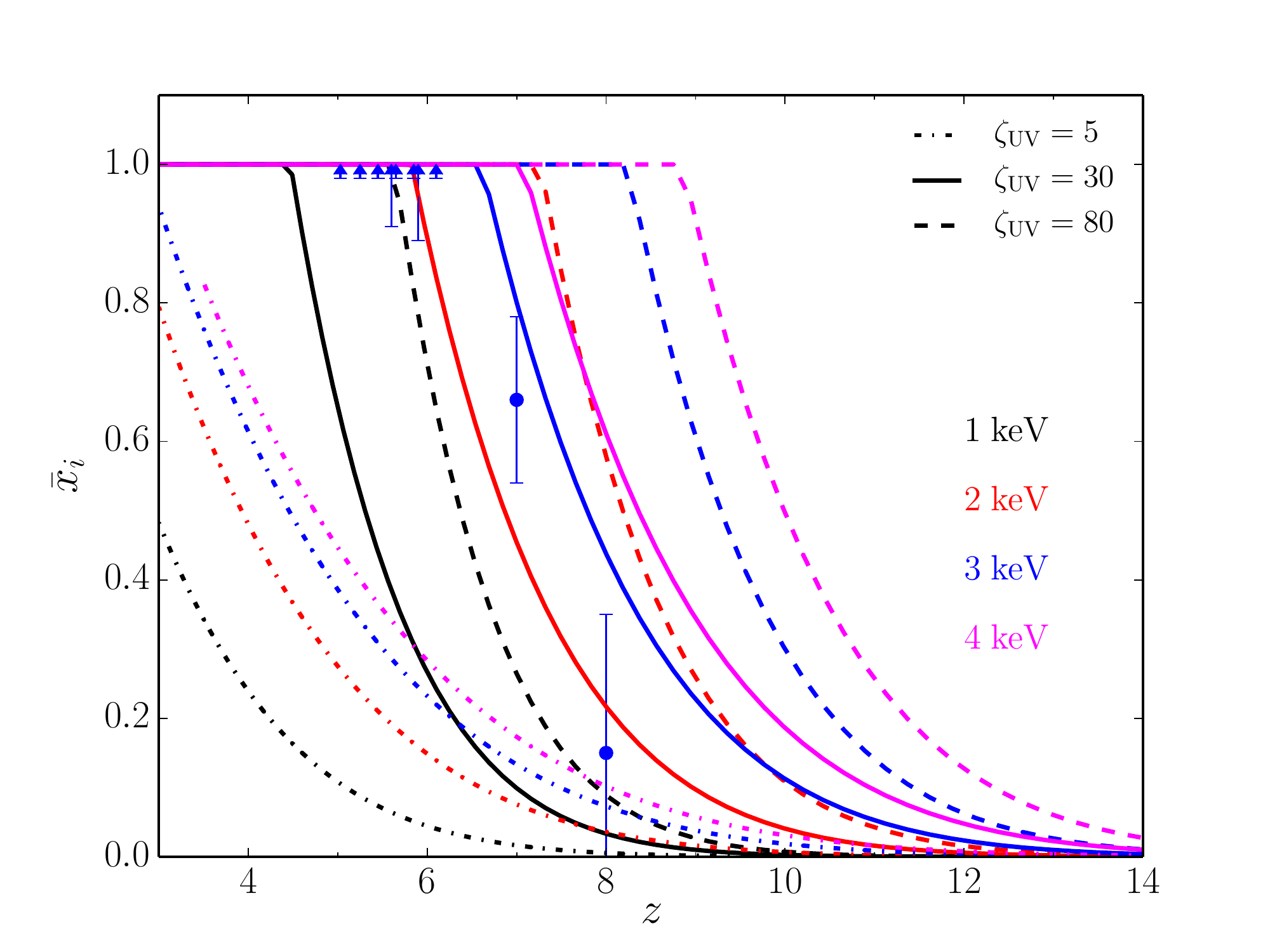} \includegraphics[width=0.49\textwidth]{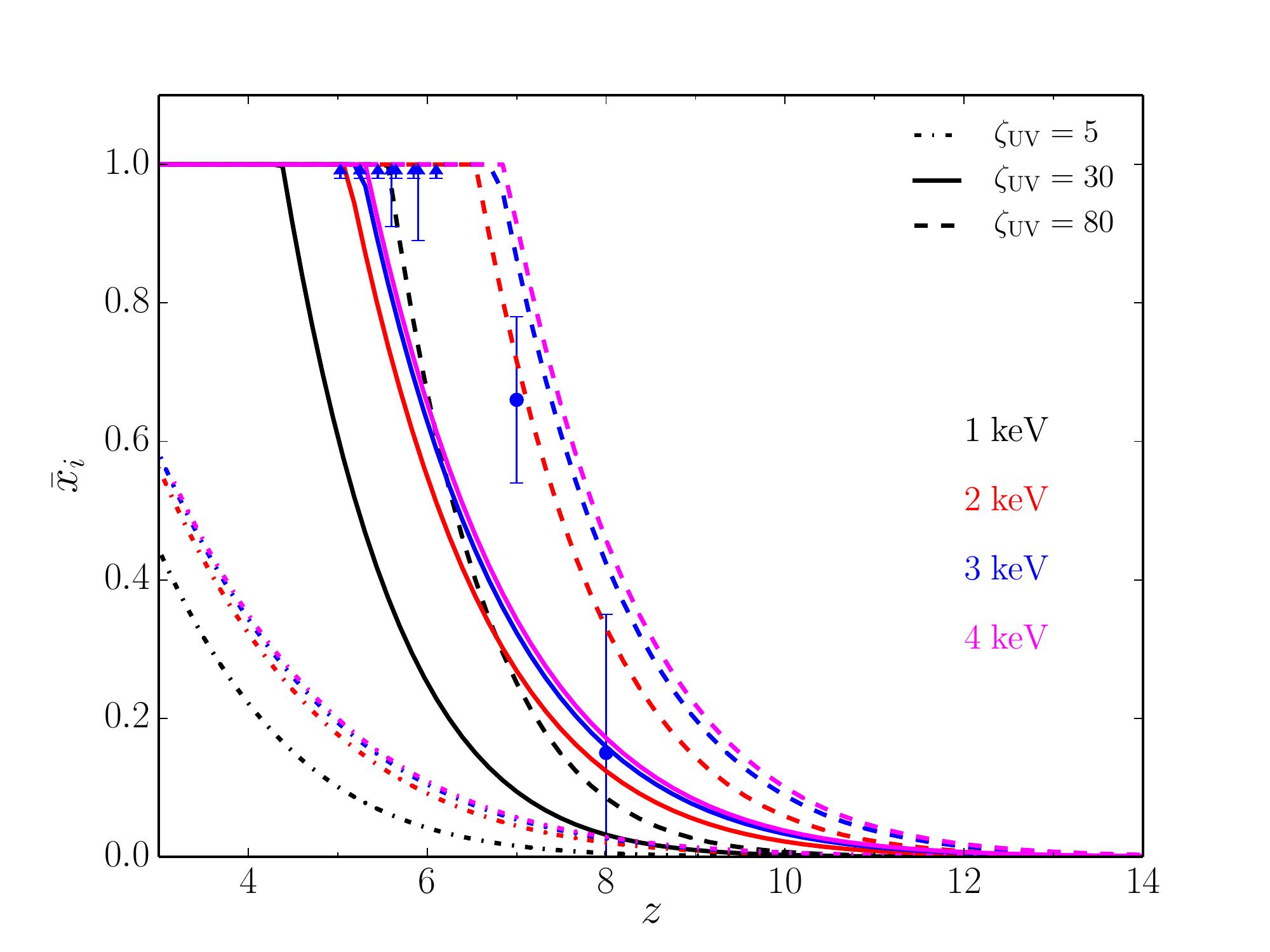}
	\end{center}
	\caption{Total ionized fraction, $\bar{x}_i$, as a function of redshift for different reionization histories, for four values of the WDM particle mass: $m_{\rm WDM} = 1$~keV (black curves), $2$~keV (red curves), $3$~keV (blue curves) and $4$~keV (magenta curves); and for three possible values of the UV ionizing efficiency: $\zeta_{\rm UV} = 5$ (dotted curves), $30$ (solid curves) and $80$ (dashed curves); fixing the minimum virial temperature to $T_{\rm vir}^{\rm min} = 10^4$~K (left panel) and $T_{\rm vir}^{\rm min} = 10^5$~K (right panel). For all cases we use $\zeta_{\rm X} = 10^{56} \, M_\odot^{-1}$. We also illustrate some of the measurements and limits currently available on $\bar{x}_i$ at a number of redshifts. See the main text for details.}
	\label{fig:xe}
\end{figure*}

\subsection{Parameter degeneracies}
\label{sec:degeneracies}

In Fig.~\ref{fig:xe} we show the total ionized fraction, $\bar{x}_i$, as a function of redshift for different reionization histories. In both panels we illustrate the redshift evolution of $\bar{x}_i$ for four values of the WDM particle mass: $m_X = 1$~keV (black curves), 2~keV (red curves), 3~keV (blue curves) and 4~keV (magenta curves); and for three values of the UV ionizing efficiency $\zeta_{\rm UV} = 5$ (dotted curves), 30 (solid curves) and 80 (dashed curves). In the left panel, the minimum virial temperature is fixed to $T_{\rm vir}^{\rm min} = 10^4$~K whereas in the right panel it is fixed to $T_{\rm vir}^{\rm min} = 10^5$~K. In both panels, we also depict some of the measurements and limits currently available on $\bar{x}_i$, that are described in the following section. Notice that, while both panels have been produced for a value of the X-ray efficiency $\zeta_{\rm X}=10^{56} \, M_\odot^{-1}$, the results are not very sensitive to this parameter, given the limited range considered here.\footnote{This is expected, as in scenarios in which the usual parameters have values close to the canonical ones, X-rays only account for a few percent of the total fraction of ionization. On the other hand, a larger contribution from X-rays can have an important impact on more extreme scenarios~\cite{Mesinger:2012ys}.} Notice also that there exists a degeneracy between $\zeta_{\rm UV}$ and $m_X$. Indeed, a lower value of the WDM mass, which implies a larger small-scale suppression and thus, a larger departure from the CDM picture, would delay structure formation and therefore the entire reionization process, and this can be compensated by a larger UV ionization efficiency. For instance, for $T_{\rm vir}^{\rm min} = 10^4$~K, the $\bar{x}_i (z)$ curves for the $m_X = 2$~keV, $\zeta_{\rm UV} = 80$ case (red dashed curve) and those for the $m_X = 4$~keV, $\zeta_{\rm UV} = 30$ case (magenta solid curve) are almost identical and they are constrained exactly in the same way from measurements and limits on $\bar{x}_i (z)$. Notice that the $\zeta_{\rm UV}$ and $m_X$ degeneracy can not broken by using the Planck results for the optical depth to reionization because $\tau$ is an integrated quantity of $\bar{x}_i$ over redshift.

Measurements of the IGM temperature $T_{\rm K} (z)$ could help in alleviating some of the degeneracies discussed above. For instance, the IGM temperature is expected to be more sensitive than $\bar{x}_i$ to $\zeta_{\rm X}$, since this is the fundamental parameter which rules the heating of the gas. However, usual temperature data are derived from the Lyman-$\alpha$ forest measurements in the fully ionized phase~\cite{Puchwein:2014zsa, Becker:2010cu}, regime where we cannot reliably compute the IGM temperature using {\tt 21cmFast}~\cite{Mesinger:2007pd, Mesinger:2010ne}. A complete and proper calculation of the IGM temperature would require computational expensive hydrodynamical simulations, which are beyond the scope of this work.

\section{Numerical analysis}
\label{sec:datanum}

\subsection{Data sets}
\label{sec:data}

The optical depth from the last-scattering surface to reionization, $\tau$, provides information of the integrated ionization history of the Universe and impacts the CMB spectrum, so that constraints on the reionization period can be obtained by means of its determination with CMB data. For our numerical analyses, we have imposed a Gaussian prior on the Planck result: $\tau = 0.055 \pm 0.009$~\cite{Aghanim:2016yuo}. We compute the redshift evolution of the total ionized fraction, $\bar{x}_i (z)$, using the {\tt 21cmFAST} code~\cite{Mesinger:2007pd, Mesinger:2010ne}, as previously explained, and in this way we determine the value of $\tau$ for each of the models studied here. In order to add the Planck prior, we have modified the CAMB (Code for Anisotropies in the Microwave Background) Boltzmann solver code~\cite{Lewis:1999bs} to allow for any possible ionization history, including those corresponding to WDM scenarios, for which we also employ the {\tt 21cmFAST} code to compute the reionization history as a function of redshift.

For deriving bounds using the ionization history of the Universe, we use some of the measurements compiled in Ref.~\cite{Bouwens:2015vha}. One set of data uses the Gunn-Peterson optical depth from bright quasars at six different redshifts, $z = 5.03, 5.25, 5.45, 4.65, 5.85, 6.10$~\cite{Fan:2005es}, whereas another one makes use of the distribution of dark gaps in quasar spectra at $z=5.6$ and $z=5.9$~\cite{McGreer:2014qwa}. Both sets of data indicate that reionization is complete by $z \sim 6$. On the other hand, the observations of Ly$\alpha$ emission in star-forming galaxies at higher redshifts ($z \gtrsim 7$), if the behavior at lower redshifts is extrapolated~\cite{Santos:2003pc, Malhotra:2004ef, McQuinn:2007dy, Mesinger:2007jr, Stark:2010qj, Stark:2010nt,  Fontana:2010ms, Dijkstra:2011de, Pentericci:2011ft, Ono:2011ew, Caruana:2012ww, Treu:2013ida, Caruana:2013qua, Tilvi:2014oia}, indicates that reionization is not complete at those high redshifts. In this work we consider recent results at $z=7$ and $z=8$~\cite{Schenker:2014tda}, which use the models of Ref.~\cite{McQuinn:2007dy}.  In practice, given the precision of our numerical simulations, the Gunn-Peterson measurements imply that reionization should be fully completed at the quoted redshifts and we take their $1\sigma$ interval as lower bounds. All these measurements of the total ionized fraction are depicted in Fig.~\ref{fig:xe}, and are included in our numerical analyses in the next section. 

Therefore, in practice, we compute two $\chi^2$, one for each type of data, and add them up. In the case of the low-redshift data of the ionization fraction, as indicated, we only consider lower bounds. In practice, for each model, at the redshifts corresponding to the data points, there is no contribution to the $\chi^2$ if $\bar{x}_i (z)$ is larger than the measured lower bound.

\subsection{Results}
\label{sec:results}

\begin{figure*}[t]
	\begin{center}
		\hspace{-5mm}
		\includegraphics[width=0.33\textwidth]{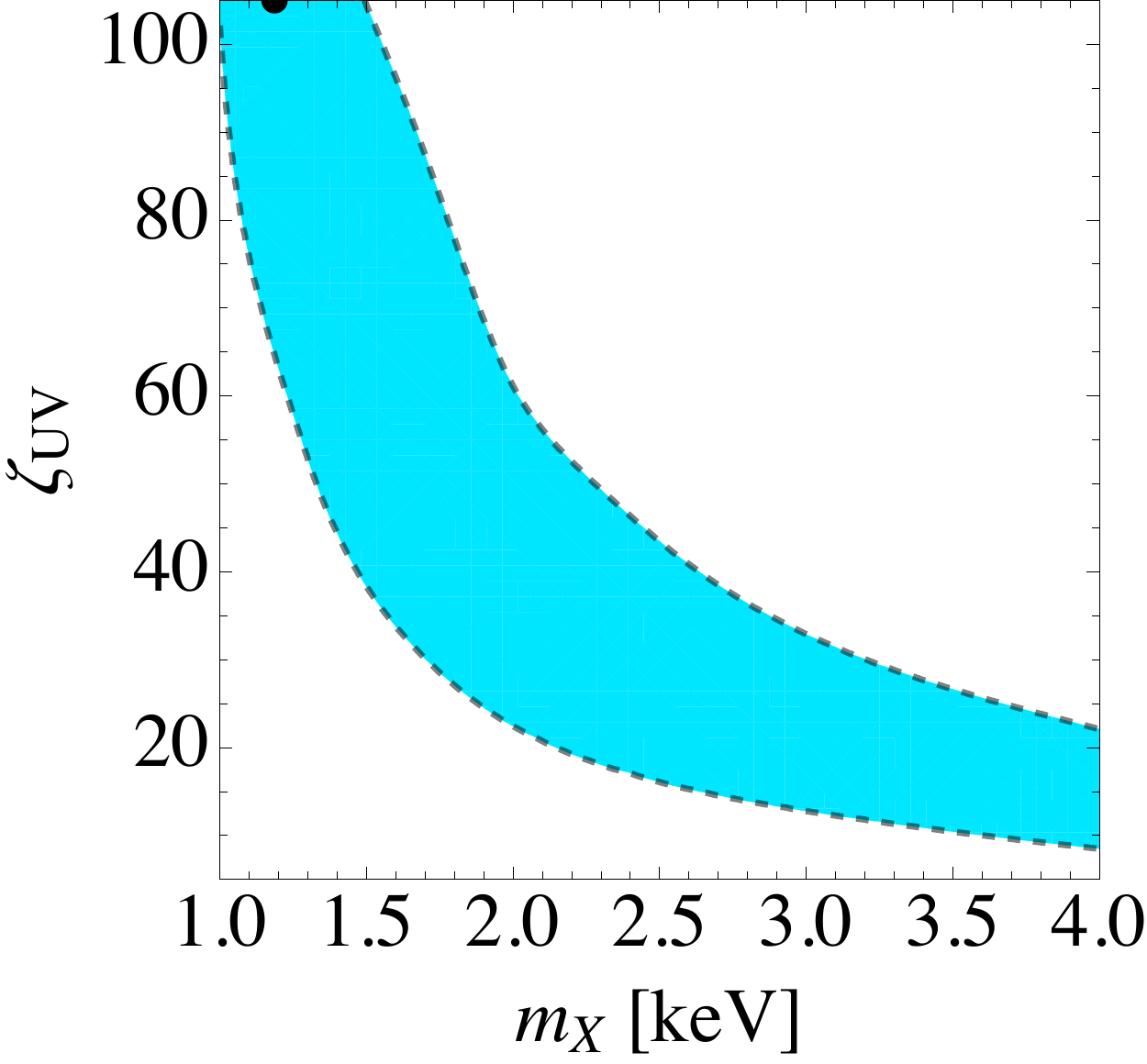}
		\includegraphics[width=0.33\textwidth]{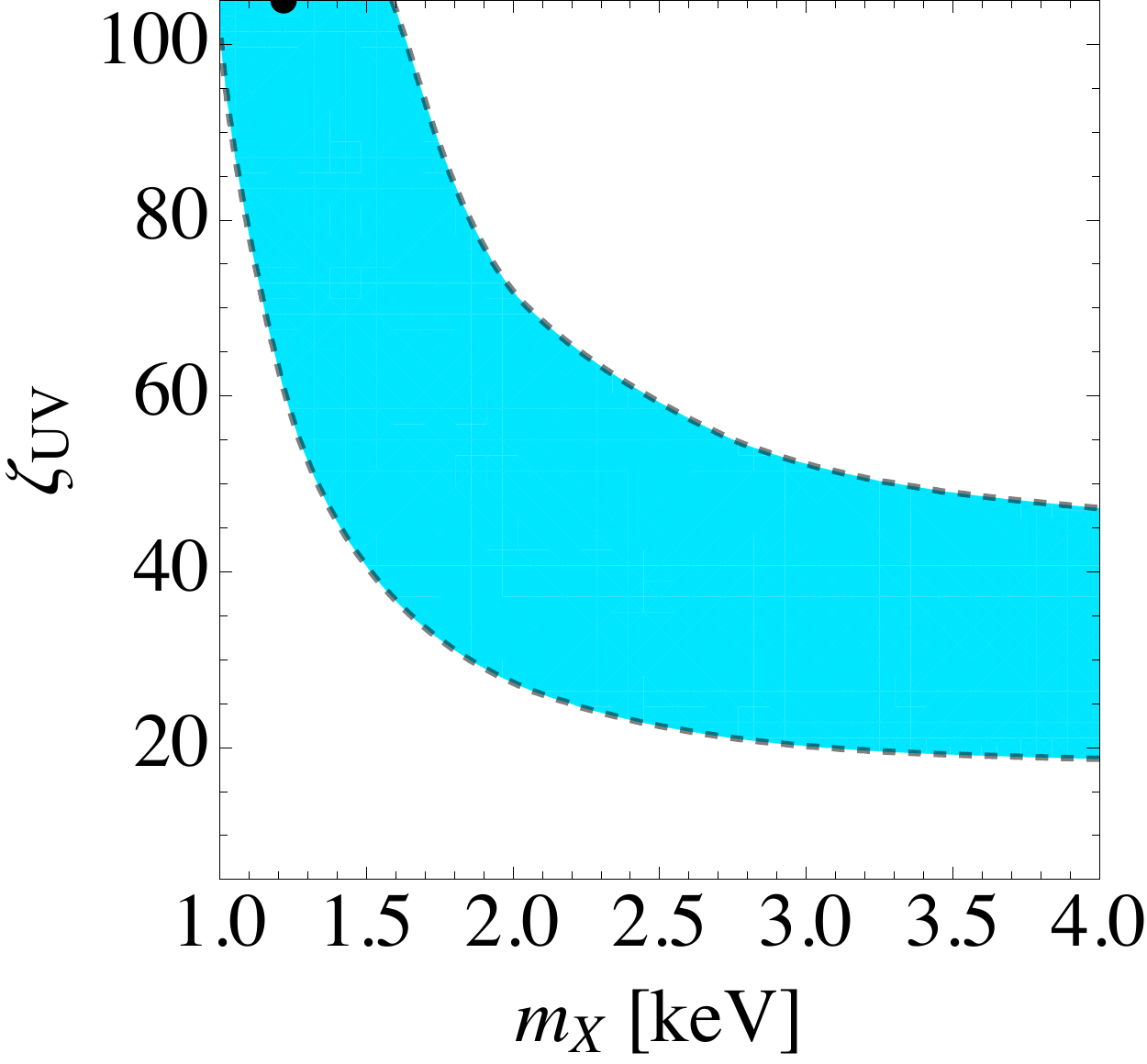} 
		\includegraphics[width=0.33\textwidth]{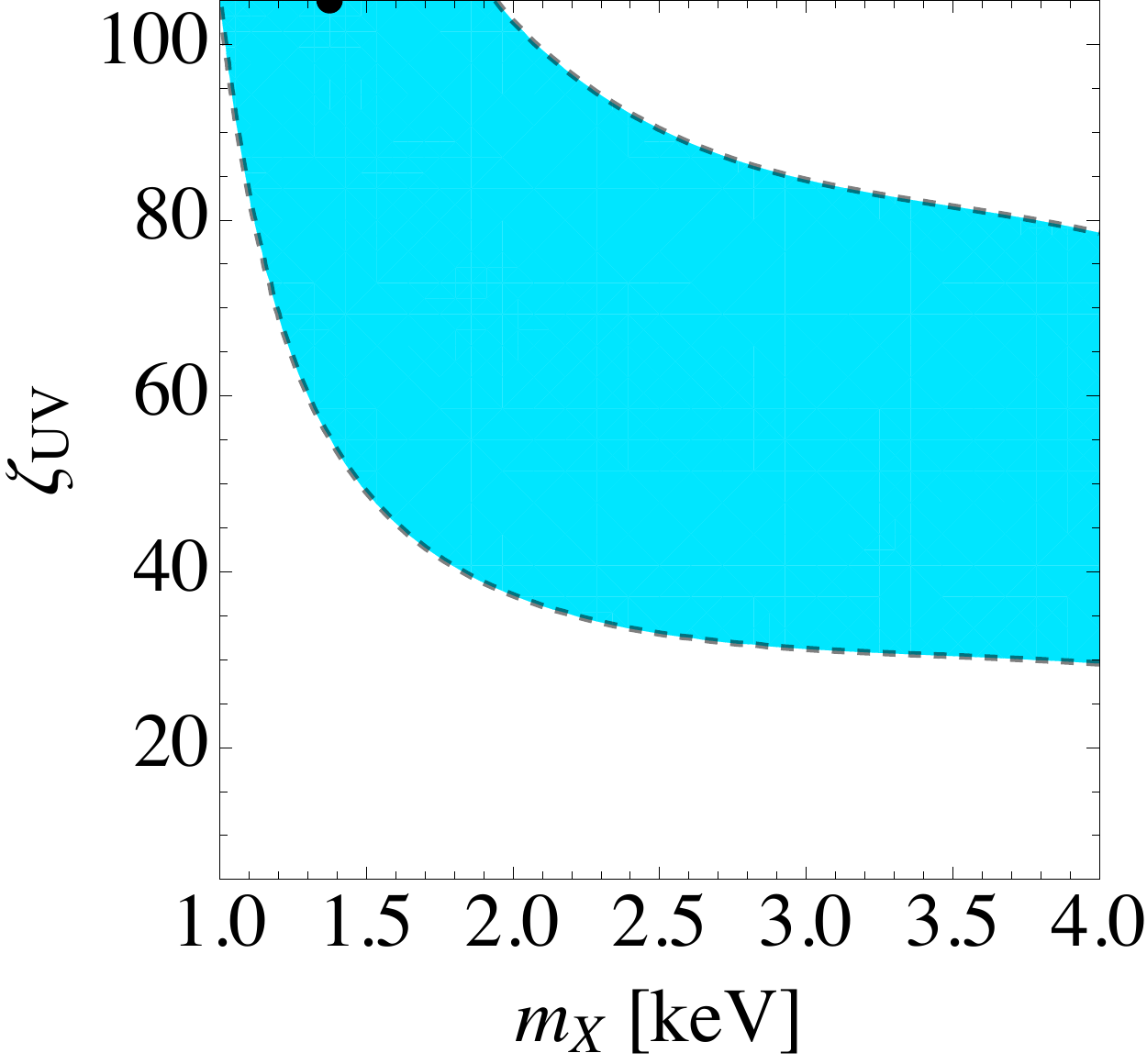} \\[2ex]
		\hspace{-5mm}
		\includegraphics[width=0.33\textwidth]{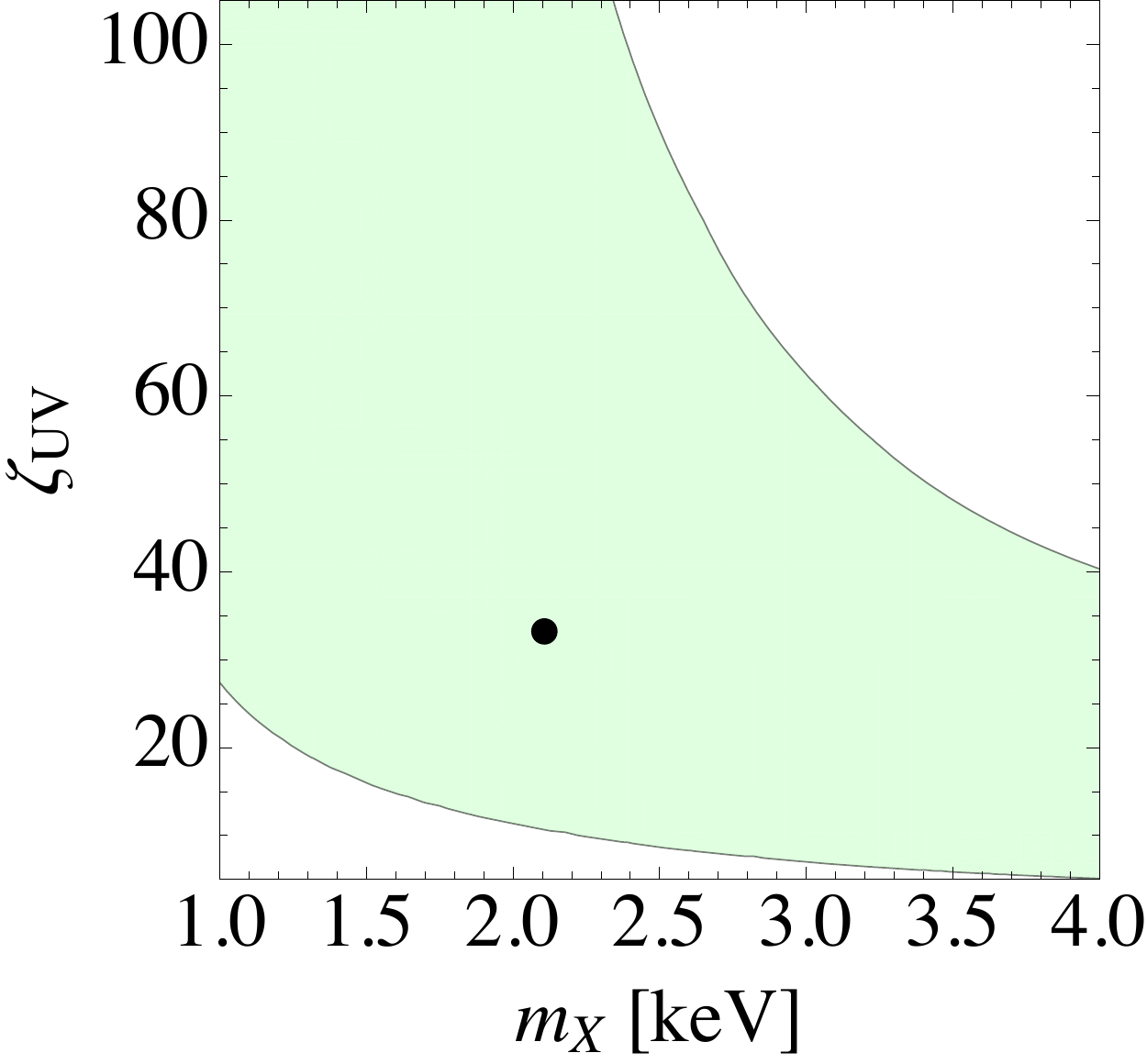}
		\includegraphics[width=0.33\textwidth]{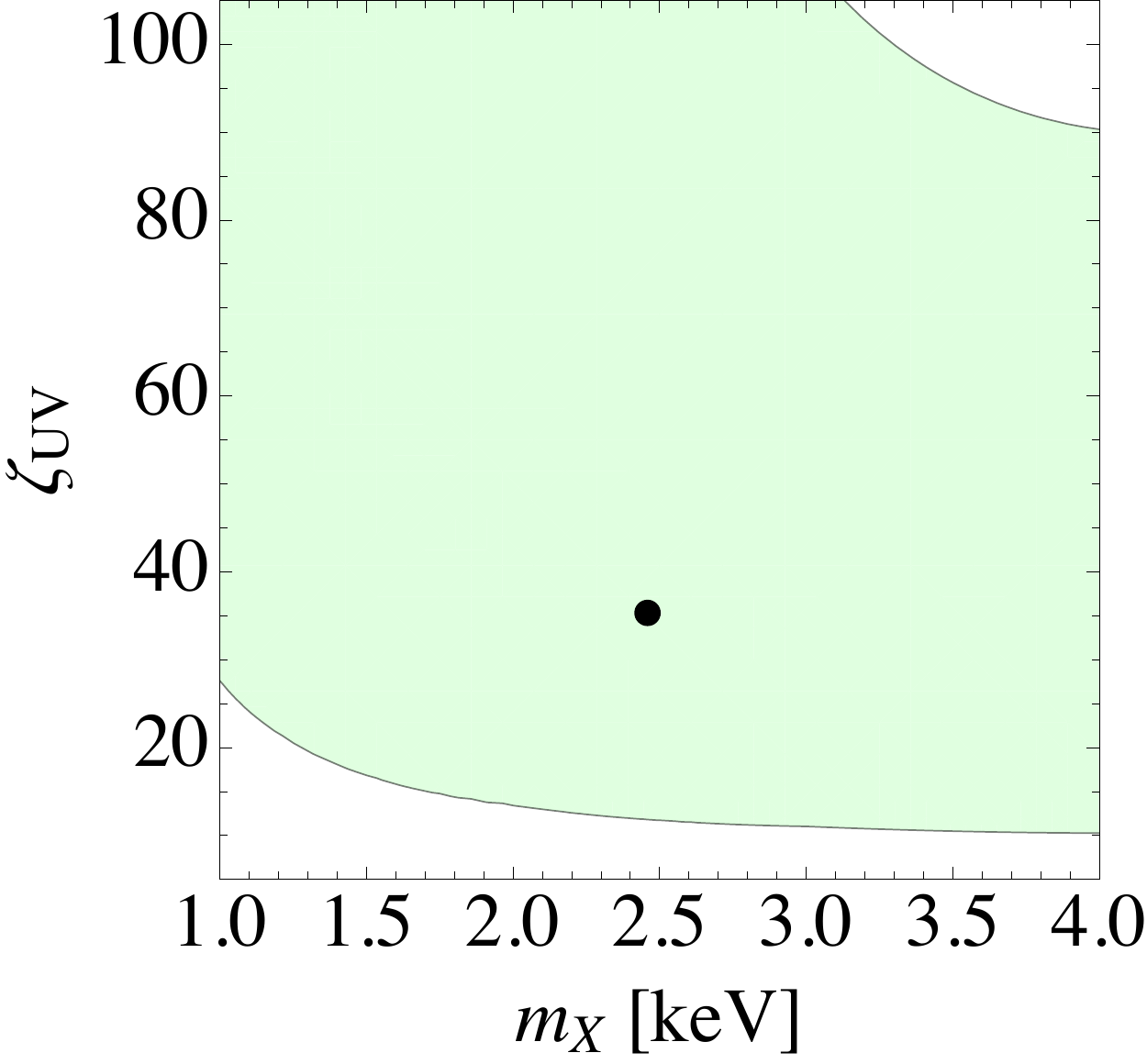} 			    \includegraphics[width=0.33\textwidth]{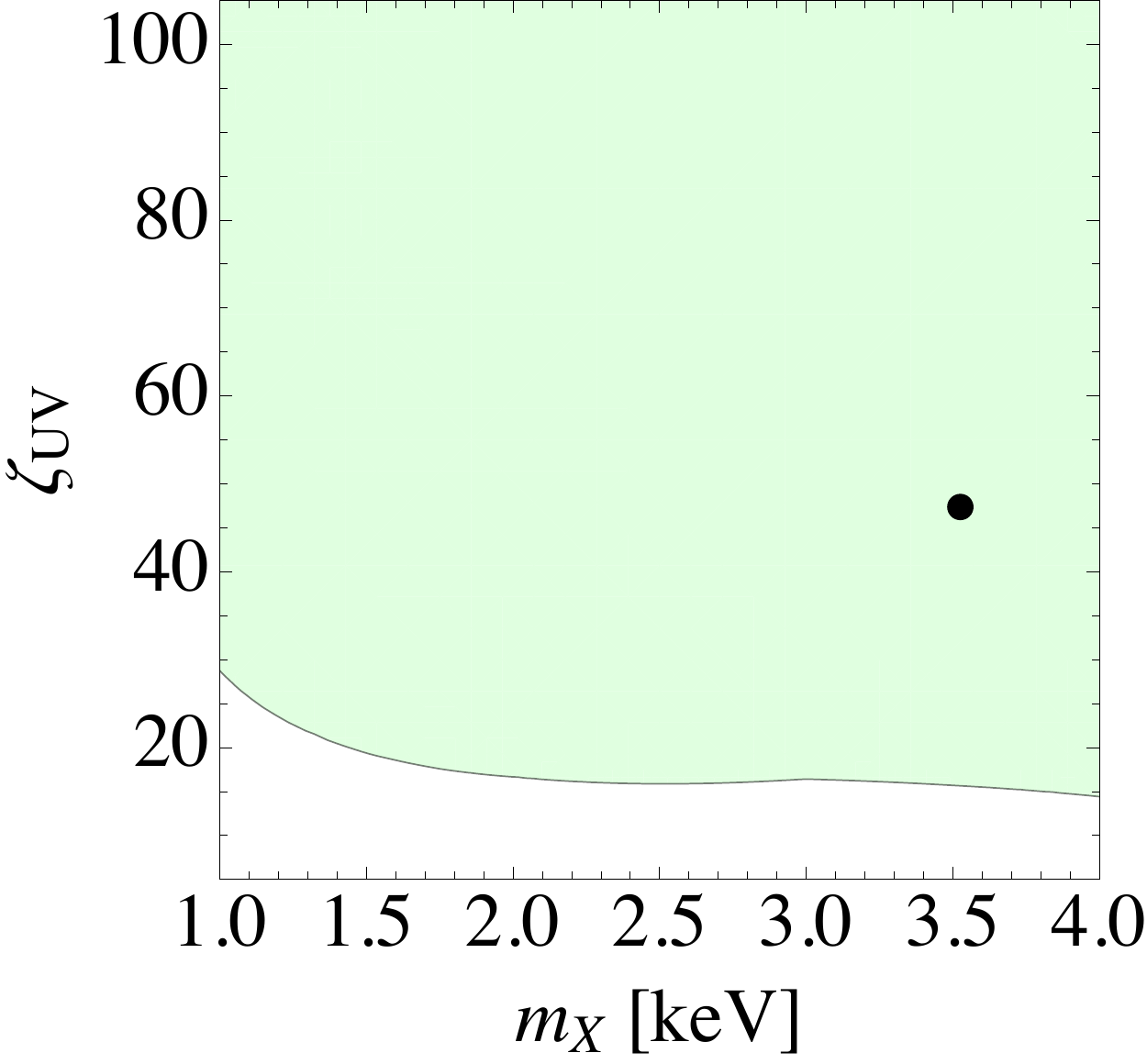}
	\end{center}
	\caption{Contours in the ($m_X$, $\zeta_{\rm UV}$) plane corresponding to 90\%~CL, for three values of the minimum virial temperature: $T_{\rm vir}^{\rm min} = 10^4$~K (left-column panels), $5 \times 10^4$~K (middle-column panels) and $10^5$~K (right-column panels). The number of X-ray photons per solar mass in stars is $\zeta_{\rm X}= 5 \times 10^{56} \, M_\odot^{-1}$ in all panels. The best-fit value is indicated with a black point. {\it Top panels}: using only $\bar{x}_i (z)$ data. {\it Bottom panels}: using only the Planck measurement of the optical depth, $\tau = 0.055 \pm 0.009$~\cite{Aghanim:2016yuo}.}
	\label{fig:chi2}
\end{figure*}

In the following, we present our results exploiting the data previously described and the simulations performed with the {\tt 21cmFAST} code. The redshift at which simulations start is $z=35$, which roughly corresponds to the epoch when the first bright sources begin to appear. The simulations stop at $z=3$, when reionization is expected to be complete.  Due to the fact the runs are computationally expensive, we have made a grid in the parameter space and we have restricted ourselves to the following values: $m_X=1,~1.5,~2,~3,~4$~keV, $\zeta_{\rm UV}= 5,~30,~42.5,~55,~80,~105$, $T_{\rm vir}^{\rm min} = 10^4$~K, $5 \times 10^4$~K and $10^5$~K, and $\zeta_{\rm X}= 10^{56} \, M_\odot^{-1}$ and $5 \times 10^{56} \, M_\odot^{-1}$. We first start describing the different constraints obtained individually by each of the data sets considered. After that, we present the combined bounds, i.e., those obtained when the measurements of the optical depth to reionization $\tau$ and $\bar{x}_i (z)$ are simultaneously considered. This approach helps to understand our final constraints and it also justifies the combination of the different data sets considered here since, as we show, there is no tension among these data sets. 

We start with the limits on our parameters obtained from measurements of $\bar{x}_i (z)$ at low redshifts. Focusing on the ($m_X$, $\zeta_{\rm UV}$) plane, we find the allowed regions to be rather independent of the adopted value of the number of X-ray photons per solar mass in stars, $\zeta_{\rm X}$. In the top panels of Fig.~\ref{fig:chi2}, we depict the regions allowed at $90\%$~CL from $\bar{x}_i (z)$ data only, corresponding to $\zeta_{\rm X} = 5 \times 10^{56}\,M_\odot^{-1}$, and three possible values of the minimum virial temperature $T_{\rm vir}^{\rm min} = 10^4$~K (left panel), $5 \times 10^4$~K (middle panel) and $10^5$~K (right panel). By comparing the results in the three panels, one can clearly see the effect of the non-trivial dependence of $\bar{x}_i (z)$ on the minimum virial temperature, as the allowed regions shift to lower values of $\zeta_{\rm UV}$ for lower values of $T_{\rm vir}^{\rm min}$.  This is because a lower minimum virial temperature imply an earlier reionization time, which then would requires a lower UV heating efficiency. The best-fit value when using this set of data is obtained for $m_X = 1.2$~keV, $\zeta_{\rm   UV} = 105$ and $T_{\rm vir}^{\rm min} = 10^4$~K, with very little sensitivity to the value of $\zeta_{\rm X}$ (the $\chi^2$ is very flat in the direction of this parameter for the range we consider).

Given that the optical depth $\tau$ only provides information on the integrated ionization history of the Universe, constraints based on its measurement are less restrictive than those obtained by using the redshift evolution of $\bar{x}_i$ (shown in the top panels of Fig.~\ref{fig:chi2}). This can be seen in the bottom panels of Fig.~\ref{fig:chi2} where we use the Planck result for $\tau$ and, analogously to the top panels, we depict the 90\%~CL allowed contour in the ($m_X$, $\zeta_{\rm UV}$) plane for $\zeta_{\rm X}= 5 \times 10^{56}\,M_\odot^{-1}$ and three values of the minimum virial temperature $T_{\rm vir}^{\rm min} = 10^4$~K (left panel), $5 \times 10^4$~K (middle panel) and $10^5$~K (right panel). As expected, qualitatively, a very similar behavior to that in the top panels of Fig.~\ref{fig:chi2} is obtained and likewise, these bounds are also insensitive to the specific value of the parameter $\zeta_{\rm X}$ within the range we consider. Quantitatively, the allowed regions are larger. In this case, the low-$m_X$ allowed contours remain basically unaffected by changes in the minimum virial temperature, as reionization is considerably delayed and high UV efficiencies are always needed, regardless of the value of $T_{\rm vir}^{\rm min}$. With this measurement, the best-fit value is obtained for $m_X = 1.6$~keV, $\zeta_{\rm UV} = 58$ and $T_{\rm vir}^{\rm min} = 2.9 \times 10^4$~K, and it is independent on the value of $\zeta_{\rm X}$ (in the range we consider for this parameter).

\begin{figure*}[t]
	\begin{center}
		\includegraphics[width=0.6\textwidth]{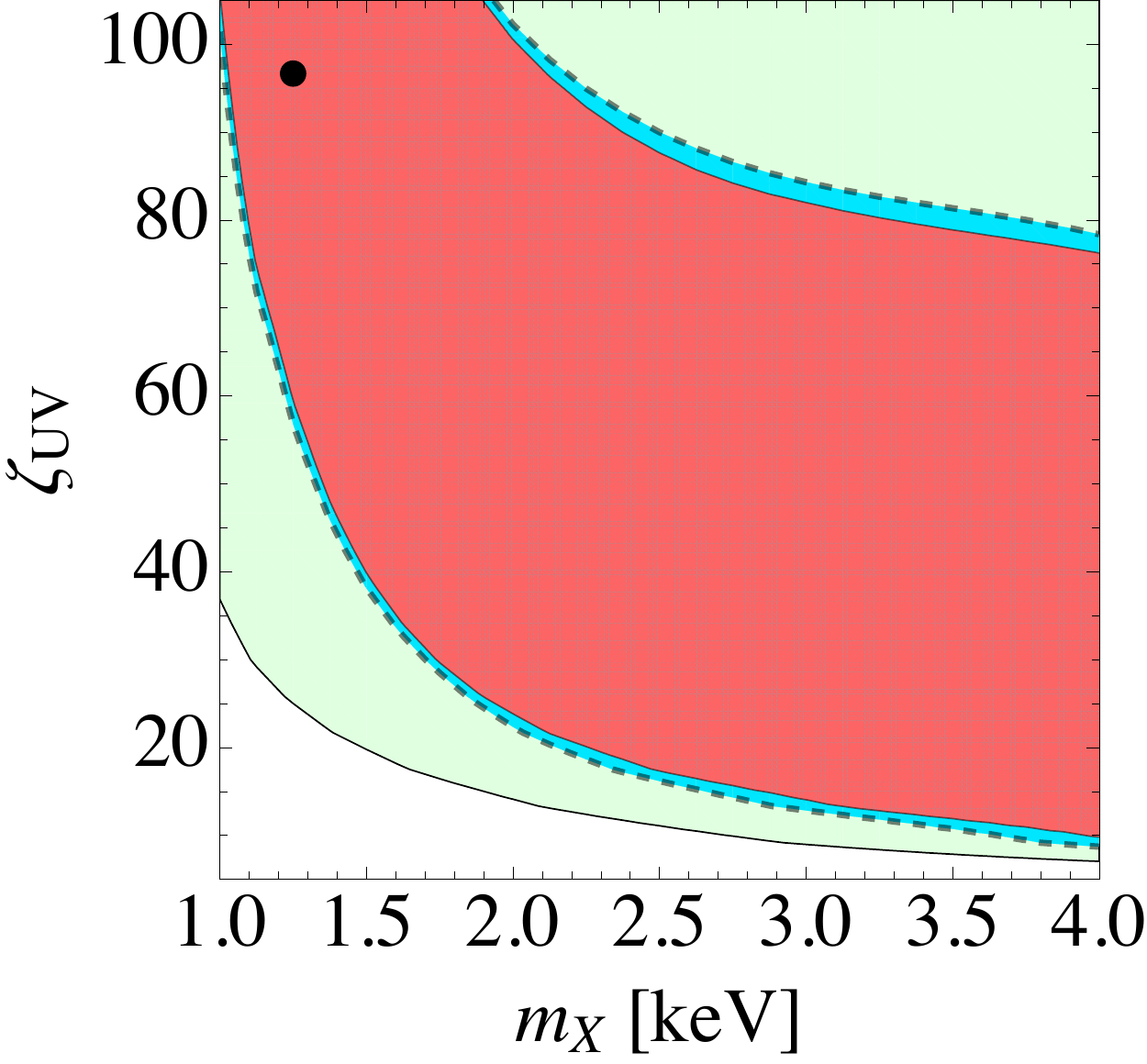}
	\end{center}
	\caption{Contours in the ($m_X$, $\zeta_{\rm UV}$) plane corresponding to 90\%~CL, for the two data sets shown in Fig.~\ref{fig:chi2}: $\bar{x}_i (z)$ (light blue region) and $\tau$ (light green region), after profiling over $T_{\rm vir}^{\rm min}$ and $\zeta_{\rm X}$, and the resulting contour obtained from the combination of them (red region). The global best-fit value in $m_X = 1.25$~keV and $\zeta_{\rm UV} = 96.6$ is indicated with a black point.}
	\label{fig:final}
\end{figure*}

In Fig.~\ref{fig:final} we show the $90\%$~CL contours in the ($m_X$, $\zeta_{\rm UV}$) plane after profiling over the minimum virial temperature $T_{\rm vir}^{\rm min}$ and the X-ray efficiency $\zeta_{\rm X}$, obtained for the two different data sets described above: the global ionization fraction $\bar{x}_i (z)$ (light blue region), and the optical depth $\tau$ (light green region), together with the final bounds after combining these data sets (red region).  Note that the $\chi^2$ is very flat along the direction of $m_X$ in the allowed region, which is an indication of the strong degeneracy between the WDM mass and other astrophysical parameters as $\zeta_{\rm UV}$ and $T_{\rm vir}^{\rm min}$. The presence of these degeneracies makes completely impossible to bound the WDM particle mass. We find the global best-fit value in $m_X = 1.25$~keV and $\zeta_{\rm UV} = 96.6$. One can check the complete consistence between both data sets, as could be expected, since the optical depth is mainly given by the integral of $\bar{x}_i (z)$ over redshift. 

\begin{figure*}[t]
	\begin{center}
		\hspace{-5mm}
		\includegraphics[width=1\textwidth]{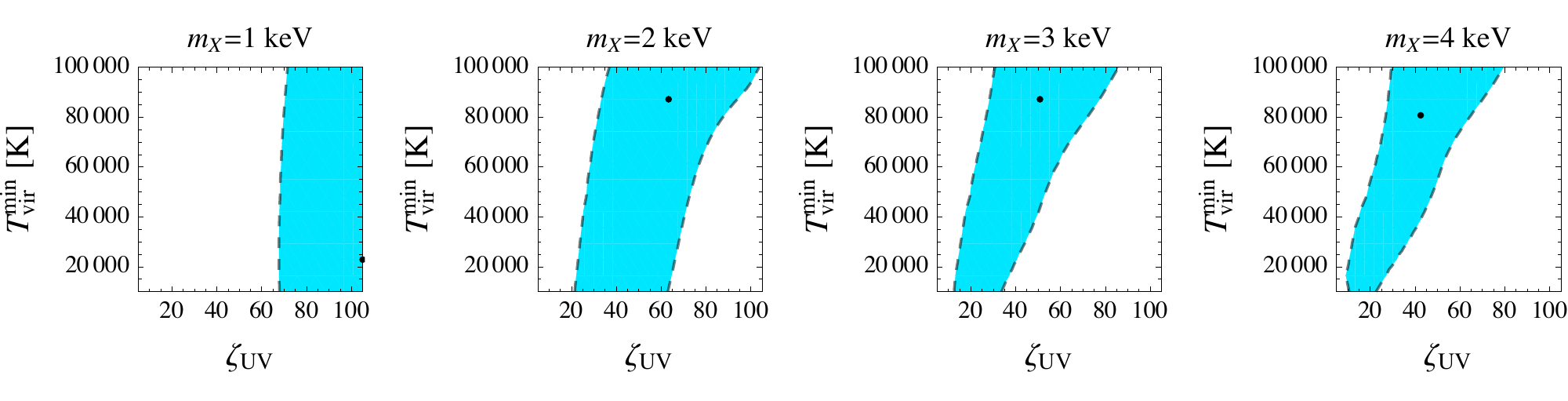}\\[2ex]
		\hspace{-5mm}
		\includegraphics[width=1\textwidth]{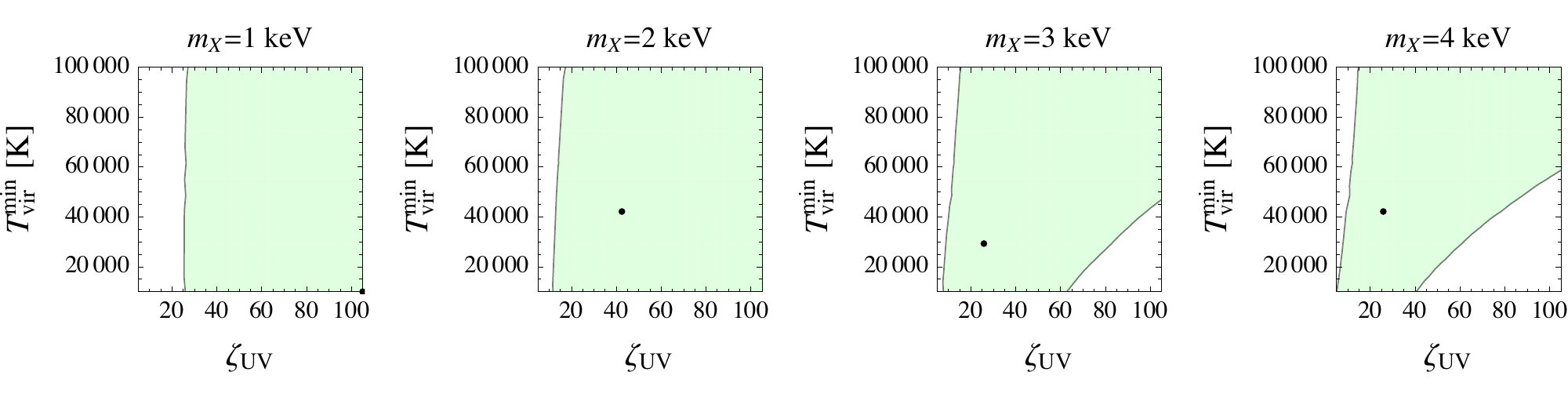}
	\end{center}
	\caption{The top (bottom) panel shows the $90\%$~CL allowed regions from the analyses to $\bar{x}_i (z)$ (optical depth, $\tau$) measurements in the ($\zeta_{\rm UV}$,$T_{\rm vir}^{\rm min}$) plane for four values of the WDM mass $m_{\rm X}$, profiling over $\zeta_{\rm X}$. The best-fit values are indicated with black points.}
	\label{fig:xi_tvir}
\end{figure*}

In order to further explore the degeneracies between the different parameters, and how the IGM quantities depend on the precise value of the WDM mass, we illustrate the $90\%$~CL contours arising from our numerical analyses in the ($\zeta_{\rm UV}$, $T_{\rm vir}^{\rm min}$) plane for different values of the WDM mass. In Fig.~\ref{fig:xi_tvir} we show the corresponding $\chi^2$ results from the fits to the ionization fraction of the universe (top panel) and of the reionization optical depth (bottom panel), profiled over $\zeta_{\rm X}$, for four different values of the WDM mass: $m_X = 1,~2,~3,$ and $4$~keV. Notice that a lower $m_X$ would imply a delayed reionization history and therefore a larger value of the UV efficiency would be required. Fig.~\ref{fig:xi_tvir} shows the strong degeneracies between these three parameters ($\zeta_{UV}$, $T_{vir}^{min}$ and $m_X$). We do not show the ($\zeta_X$, $T_{vir}^{min}$) and ($\zeta_{UV}$, $\zeta_X$) planes, because, as he have pointed out earlier, $\bar{x}_i$ and $\tau$ are almost insensitive to changes in $\zeta_X$ for the range of interest.

Let us finally comment that measurements of the Ly$\alpha$ power spectrum can be also used to constrain simultaneously the WDM mass and the IGM thermal history,  inferred from the suppression of power in the matter spectrum and via Jeans and Doppler broadening of the absorption lines~\cite{Gnedin:1997td, Theuns:1999mz}, respectively. By means of this method the density of neutral hydrogen can be estimated and then used to extract the total matter density. Indeed, for a given DM scenario, the recently observed cutoff in the Ly$\alpha$ flux power spectrum can be related to the IGM thermal history~\cite{Viel:2013apy, Garzilli:2015iwa} via a temperature-density relation~\cite{Hui:1997dp}, obtaining more constraining results than the ones presented here~\cite{Viel:2005qj, Seljak:2006qw, Viel:2006kd, Viel:2007mv, Boyarsky:2008xj, Viel:2013apy, Baur:2015jsy, Irsic:2017ixq, Yeche:2017upn}.

\pagebreak
\section{Conclusions}
\label{sec:conclusions}

Our knowledge of the reionization of the Universe is still obscure. Measurements of the CMB provide information on the so-called optical depth to reionization. The most recent analysis from the Planck collaboration using the Planck HFI obtained a relatively low value, $\tau = 0.055 \pm 0.009$~\cite{Aghanim:2016yuo}, based exclusively on the spectrum polarization data. Additional (non-integrated) information on the ionization history of the Universe can also be extracted from measurements of the Gunn-Peterson optical depth or of the damping absorption wings in bright quasar spectra, or from the prevalence of Ly$\alpha$ emission in the spectra of high-redshift galaxies. In addition, measurements of the thermal history, sensitive to the ionization history, are also important.

Another crucial ingredient to understand the formation of structure at late times is the DM nature. While CDM models can satisfactorily explain large scale structure observations, there are a number of discrepancies at small scales between observations and CDM predictions. This small-scale crisis of the CDM paradigm could be alleviated in WDM scenarios for which the associated free-streaming length is larger and thus, fluctuations at small scales would be suppressed, potentially explaining the missing satellite~\cite{Klypin:1999uc, Moore:1999nt}, too-big-to-fail~\cite{BoylanKolchin:2011dk} and the core-cusp~\cite{Moore:1999gc, Springel:2008cc} problems. Reionization in WDM cosmologies would be delayed for this very same effect: halo formation processes would be delayed, and so would the onset of reionization. However, the details of reionization are not known accurately and precise determinations of the number of X-ray photons per solar mass in stars, the UV ionization efficiency and the minimum virial temperature that sets the threshold mass for halos to host star-forming galaxies are only moderately constrained. 

In this work, we have considered measurements of the reionization optical depth and the Universe's ionized fraction to constrain the mass of the DM particle in WDM scenarios. Using the {\tt 21cmFAST} code~\cite{Mesinger:2007pd, Mesinger:2010ne}, we have performed simulations of different ionization histories in WDM cosmologies, including their corresponding halo mass function, discussed in Sec.~\ref{sec:HMF}. As described in Sec.~\ref{sec:reioWDM}, we have considered four free parameters: the WDM mass, the number of X-ray photons per solar mass in stars, the minimum virial temperature for halos to host galaxies and the UV ionization efficiency. However, there are important degeneracies among these parameters. For instance, the lower the WDM mass the larger the free-streaming length and thus, the longer structure formation delays, which in turn can be compensated by a larger UV ionizing efficiency or a smaller minimum virial temperature. Therefore, given the degeneracies among some of the astrophysical parameters that drive the ionization processes and the WDM mass, i.e., the suppression of the matter power spectrum, obtaining constraints on the minimum mass of the WDM particle with this approach is a difficult task, and therefore our results can not compete to those obtained from measurements of the IGM temperature in the post-reionization era~\cite{Viel:2005qj, Seljak:2006qw, Viel:2006kd, Viel:2007mv, Boyarsky:2008xj, Viel:2013apy, Baur:2015jsy, Irsic:2017ixq, Yeche:2017upn}.

Finally, let us stress that future measurements of the 21~cm hyperfine transition of neutral hydrogen, which would map its distribution at different redshifts (and thus the distribution of $\bar{x}_i (z)$), are expected to constitute a very useful tool to understand the ionization history of the Universe, and could allow to further test predictions from WDM models~\cite{Loeb:2003ya, Mesinger:2013nua, Sitwell:2013fpa, Sekiguchi:2014wfa, Shimabukuro:2014ava, Carucci:2015bra} or even to disentangle the potential signals from DM annihilations or decays in CDM scenarios~\cite{Shchekinov:2006eb, Furlanetto:2006wp, Valdes:2007cu, Chuzhoy:2007fg, Cumberbatch:2008rh, Natarajan:2009bm, Yuan:2009xq, Valdes:2012zv, Evoli:2014pva, Lopez-Honorez:2016sur, Poulin:2016anj}.

\section*{Acknowledgments}
We thank A. Mesinger and J. Miralda-Escud\'e for clarifications and enlightening discussions. We also thank A. C. Vincent, who took part on the initial stages of this work, for useful comments. LLH is supported by the FNRS-FRS and also acknowledges partial support by the Universit\'e Libre de Bruxelles, the Vrije Universiteit Brussel (VUB), the Belgian Federal Science Policy Office through the Interuniversity Attraction Pole P7/37, the IISN and the Strategic Research Program \textit{High-Energy Physics} of the VUB. OM and PVD are supported by PROMETEO II/2014/050 and by the Spanish Grant FPA2014--57816-P of the MINECO. SPR is supported by a Ram\'on y Cajal contract, by the Spanish MINECO under grant FPA2014-54459-P and by the Generalitat Valenciana under grant PROMETEOII/2014/049. OM, SPR and PVD are also supported by the MINECO Grant SEV-2014-0398 and by the European Union's Horizon 2020 research and innovation program under the Marie Sk\l odowska-Curie grant agreements No. 690575 and 674896. SPR is also partially supported by the Portuguese FCT through the CFTP-FCT Unit 777 (PEst-OE/FIS/UI0777/2013).

\bibliographystyle{apsrev4-1}
\bibliography{bib21}

\end{document}